\documentclass[sigconf]{acmart}

\usepackage{booktabs} % For formal tables

\usepackage{amssymb}
\usepackage{amsmath}
\usepackage{amscd,amsfonts,amsbsy,rotating}
\usepackage{balance}
\usepackage{graphicx}
\usepackage{epsfig,epstopdf}
\usepackage{subfigure}
\usepackage{multirow}
\usepackage{booktabs}
\usepackage{color,xcolor}
\usepackage{url}
\usepackage{latexsym,bm}
\usepackage{enumitem,balance,mathtools}
\usepackage{wrapfig}
\usepackage{algorithm}
\usepackage{algorithmic}
\usepackage{ifpdf}
\usepackage{diagbox}
\usepackage{caption}
\usepackage{makecell}
\usepackage{subfigure}
\usepackage{bm}
\usepackage{hyperref}

\newcommand{\bs}{\boldsymbol}
\newcommand{\bc}{\bs{c}}
\newcommand{\bx}{\bs{x}}
\newcommand{\bh}{\bs{h}}
\newcommand{\bss}{\bs{s}}

\usepackage{array}
\newcolumntype{N}{@{}m{0pt}@{}}

\newcommand{\minisection}[1]{\vspace{5pt}\noindent\textbf{#1.}}

\begin{document}

\copyrightyear{2018} 
\acmYear{2018} 
\setcopyright{acmcopyright}
\acmConference[CIKM '18]{The 27th ACM International Conference on Information and Knowledge Management}{October 22--26, 2018}{Torino, Italy}
\acmBooktitle{The 27th ACM International Conference on Information and Knowledge Management (CIKM '18), October 22--26, 2018, Torino, Italy}
\acmPrice{15.00}
\acmDOI{10.1145/3269206.3271677}
\acmISBN{978-1-4503-6014-2/18/10}

\fancyhead{}

\title{Learning Multi-touch Conversion Attribution \\with Dual-attention Mechanisms for Online Advertising}

 \author{Kan Ren, Yuchen Fang, Weinan Zhang, Shuhao Liu, \\
 	Jiajun Li, Ya Zhang, Yong Yu, Jun Wang
 }
 \affiliation{%
 	\institution{Shanghai Jiao Tong University, University College London \\
 		\{kren, arthur\_fyc, wnzhang, yyu\}@apex.sjtu.edu.cn, ~~ j.wang@cs.ucl.ac.uk
 }
 }

\begin{abstract}
In online advertising, the Internet users may be exposed to a sequence of different ad campaigns, i.e., display ads, search, or referrals from multiple channels, before led up to any final sales conversion and transaction. For both campaigners and publishers, it is fundamentally critical to estimate the contribution from ad campaign touch-points during the customer journey (conversion funnel) and assign the right credit to the right ad exposure accordingly. However, the existing research on the multi-touch attribution problem lacks a principled way of utilizing the users' pre-conversion actions (i.e., clicks), and quite often fails to model the sequential patterns among the touch points from a user's behavior data. To make it worse, the current industry practice is merely employing a set of arbitrary rules as the attribution model, e.g., the popular \emph{last-touch} model assigns 100\% credit to the final touch-point regardless of actual attributions. In this paper, we propose a Dual-attention Recurrent Neural Network (DARNN) for the multi-touch attribution problem. It \textit{learns} the attribution values through an attention mechanism directly from the conversion estimation objective. To achieve this, we utilize sequence-to-sequence prediction for user clicks, and combine both post-view and post-click attribution patterns together for the final conversion estimation. To quantitatively benchmark attribution models, we also propose a novel yet practical attribution evaluation scheme through the proxy of budget allocation (under the estimated attributions) over ad channels. The experimental results on two real datasets demonstrate the significant performance gains of our attribution model against the state of the art.
\end{abstract}


% \settopmatter{printacmref=false} % Removes citation information below abstract

\maketitle

\section{Introduction}\label{sec:intro}

% online advertising

A  benefit for online advertising is that advertisers would be able to get a significant amount of user feedbacks to measure the successfulness of their ad campaigns and optimize them accordingly. Aiming at delivering the optimization above, computational advertising has gained a large attraction and achieved great progress in many technical fields, including user targeting \cite{mcmahan2013ad,ren2016user}, bidding strategy \cite{perlich2012bid,zhang2014optimal,ren2017bidding} and budget pacing \cite{amin2012budget,agarwal2014budget,lee2013real}.

\begin{figure}[t]
	\centering
	\includegraphics[width=0.8\columnwidth]{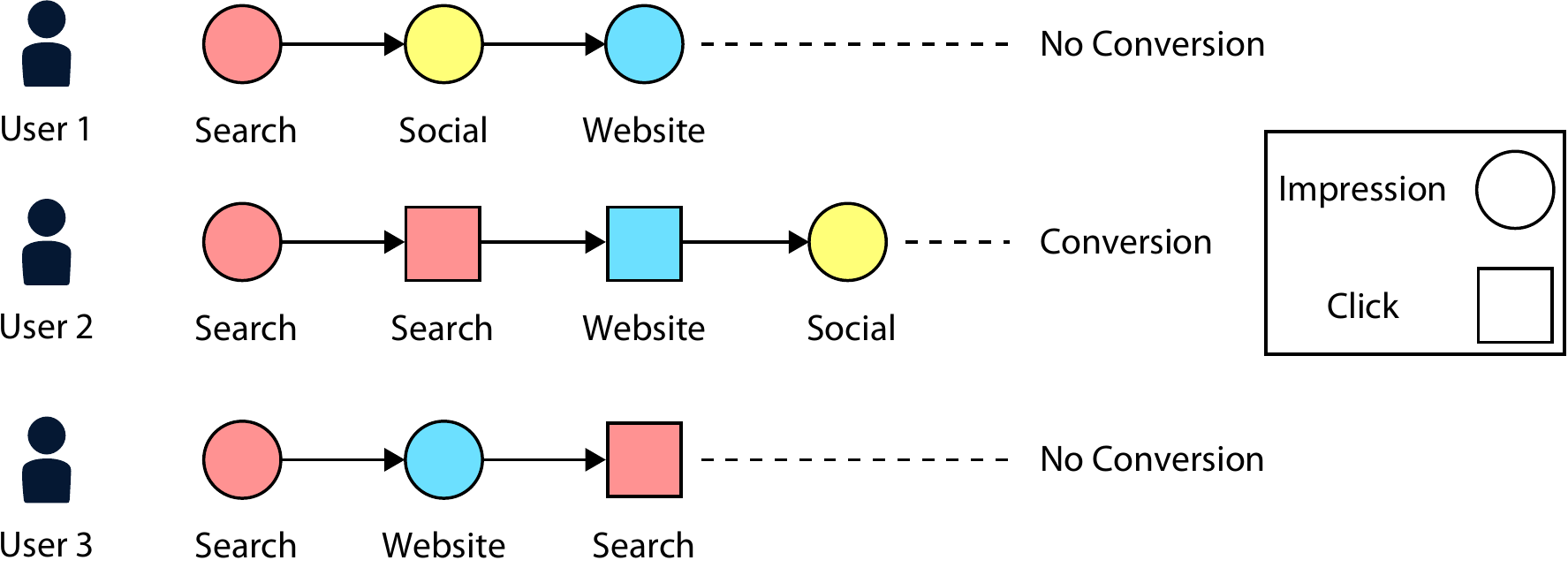}
	\caption{An illustration of different user activity sequences over multiple channels. Here we illustrate two user interactions with the ad contents, i.e., impressions and clicks, over three typical ad channels. Each user would gain an impression and then would probably click on that. After a sequence of user actions, the final conversion may be drawn according to the comprehensive experiments.}\label{fig:hori-vert-ads}
\end{figure}

As illustrated in Figure~\ref{fig:hori-vert-ads}, with online advertising, an Internet user may be exposed to a sequence of ad campaigns from  multiple channels, such as search engines, social media, mobile platforms, before reaching to any final conversion and transaction. 
It is crucial for advertisers to attribute the right conversion credit onto each \emph{touch point} (i.e., the interaction between the user and the ad content, along the customer journey).
The reasons are threefold.
First, advertisers should know the contribution of each single touch point to the final conversion so as to make informed impression-level ad buying decisions \cite{lee2012estimating}.
Second, if the attribution of each conversion over multiple ad exposures can be accurately and reliably estimated, a more quantitative credit-based ad pricing scheme can be established between advertisers and publishers (and ad tech providers). Last but not least, the attribution aggregated over ad channels may provide useful guidance for advertisers to allocate their budgets over these ad channels so as to acquire more positive user actions with lower cost in the next-round campaigns \cite{geyik2014multi}.

% traditional attribution model
% several categories: 1. rule-based; 2. simple data-driven model; 3. additive exciting process; 4. sequential process (Markov process)

Traditionally, the attribution problem is addressed simply  by a rule-based approach among most advertisers, such as first touch, last touch and other simple mechanisms \cite{wang2017display}. Specifically, the first (last) touch method generally attributes the conversion credits to the first (last) user interaction with the ad content. While these methods are easy to deploy, they obviously lack adequate capability of useful pattern recognition to support the higher-level budget allocation \cite{chandler2009measuring}.
\citet{shao2011data} proposed the first data-driven multi-touch attribution model to allocate the credits to all the user touch points.
Whereafter, many works have been published including probabilistic models using some distributional assumptions \cite{xu2016lift,dalessandro2012causally} and additive exciting process \cite{zhang2014multi,ji2017additional,xu2014path}.
Despite of the claimed advantages, there are three important factors missing in the above solutions.

\textbf{Sequential Pattern Modeling.}
These methods are all based on the assumption that the user conversion would be driven by the individual advertising touch point of positive influence \cite{zhang2014multi,shao2011data}, which may not be realistic for the user journey (conversion funnel).
In fact, sequential patterns within the user browsing behavior are of great value for response prediction or decision making in many fields such as recommender systems \cite{rendle2010factorizing}, information retrieval \cite{song2017hierarchical} and search advertising \cite{zhang2014sequential}.

% drawbacks:
% 	1. heuristic assumption about the attribution mechanism
% 	2. not learning about the attritbution
% 	3. ignore the difference between user behaviors

\begin{figure}[t]
	\centering
	\includegraphics[width=1.0\columnwidth]{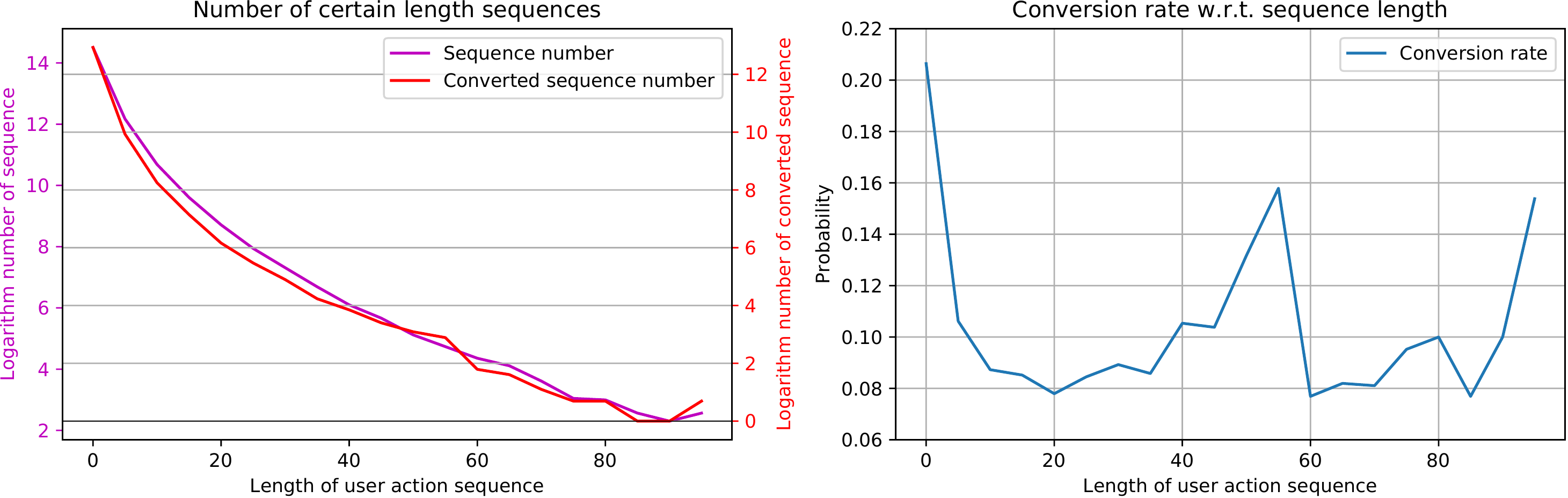}
	\caption{The statistics of action sequence lengths and the conversion rates over Criteo dataset. The left plot shows the number of sequences and the converted sequences w.r.t. sequence length; the right plot presents the user conversion rate w.r.t. the user action sequence length.}\label{fig:user-sequence}
\end{figure}

\textbf{Data-driven Credit Assignment.} The attribution credits obtained in these models are heuristically assigned to each user interaction with the advertiser's contents, rather than statistically learned from the data.
For example, \citet{ji2017additional} proposed that the final conversion would be driven by the additive hazard rate of being converted at the time of each previous touch point, which pre-assumes that the more user exposures, the higher probability of the final user action.
As is illustrated in Figure~\ref{fig:user-sequence} of one real-world dataset used in our experiments, the conversion rate does not necessarily increase w.r.t. the user action sequence length.
This assumption may cause unconscionable ad exposures and may destruct user experience in online service.

\textbf{Different Pre-conversion Behaviors.} Almost all the related works ignore the difference between various types of user behaviors.
Specifically, they assume the attribution are solely based on post-view or post-click, or even simply treat these behavior types equally for conversion attribution, where the credits are placed solely on impressions or clicks, or even discard the difference between them.
These treatments are not effective since the user shows apparently different preferences behind the different interactive actions, which may (not) lead to the final conversion in different degrees.

% attention as attribution
To address the above limitations, in this paper, we propose a Dual-attention Recurrent Neural Network (DARNN) to capture the sequential user behavior patterns and learn the optimal attentions as the conversion attributions.
Specificall, our model has two learning objectives.
On one hand, we utilize sequence-to-sequence architecture to model the relationship between the impressions and the click actions, where the click behavior modeling is handled in this procedure.
On the other hand, the final sequence prediction is the probability of the user conversion with the attention learned from the sequential modeling.
The advantage of the attention mechanism is that it not only contributes to the sequence prediction accuracy, but also naturally \textit{learns} the attribution of the conversion action over the whole sequence of the touch points.
Moreover, DARNN applies the attention mechanism not merely on the features of the original touch point, but additionally over the learned hidden states of click actions, and then dynamically combines both attentions to predict the final conversion, which is the reason of \textit{dual} attention.
By this means the conversion estimation has captured both impression-level and click-level patterns.
We also note that both dual-attention and the dynamic combination for the final conversion prediction are statistically learned from the data.

% attribution could be applied backward to direct the budget allocation.
In addition, we also propose an offline evaluation framework for conversion attribution mechanisms.
Since the obtained attribution credits over different channels could direct the budget allocation for the subsequent ad campaigns.
However, none of the related work has empirically shown the effectiveness of the calculated attributions \cite{shao2011data,dalessandro2012causally,zhang2014multi,ji2016probabilistic,ji2017additional,xu2014path} unless spending a huge budget to conduct online A/B test \cite{xu2016lift,geyik2014multi}.
Since budget allocation over different ad channels is always an important decision for advertisers to make, it is crucial for them to evaluate the performance of their multi-touch attribution methods, before the online A/B testing phase.

% novelty
To sum up, the novelty of our work is fourfold. (i) We build sequential pattern learning models for user behavior sequence. (ii) Our model learns the attribution from the final conversion estimation rather than heuristically assigning credits. (iii) We combine different attributions on various user action types. (iv) We also propose an offline evaluation protocol to measure the effectiveness of the attribution model through ad budget allocation and campaign data replay.
Our experimental results prove the significant improvement (over 5.5\%) of our DARNN model on conversion estimation performance against state-of-the-art baselines.
The back replay evaluation also illustrates that the proposed model achieves the best cost effective performance.

\section{Related Works}\label{sec:related-work}
% performance-based advertising
In online advertising, conversion attribution is commonly calculated by some rule-based methods, such as first-touch and last-touch, whereafter the return-on-investment (ROI) is gained based on the achieved attribution results which may result in some bias \cite{chandler2009measuring}.
In recent years, many works based on multi-touch attribution (MTA) have been proposed for modeling the attribution for the sequential touch points over various channels \cite{berman2017beyond,sinha2014estimating}.
\citet{shao2011data} proposed the first work for data-driven multi-touch attribution model, which estimates the conversion rate based on the viewed ads of the user by the bagged logistic regression model.
Some other works are mainly based on the probabilistic models with some distributional assumptions.
\citet{dalessandro2012causally} proposed a causally motivated methodology that the conversion credits should be assigned by a causal cooperation model such as Shapley value.
\citet{xu2016lift} argued that the user behavior has different additional effect on the final user decision of conversion and proposed a lift-based prediction model for real-time ad delivery.
However, these methods did not take all the touch points of a user into the whole consideration so that the temporal and sequential factors were ignored \cite{ji2017additional}.
Moreover, these models did either not consider much of sequential pattern modeling, which has been shown great effectiveness of user modeling \cite{zhang2014sequential}.

% point process model
As for the multiple interactions between advertisers and users, many works proposed the exciting point process methods for user behavior modeling.
\citet{yan2015machine} developed a two-dimensional Hawkes process model to capture the influences from sellers' activities on their contributions to the winning outcome in sales pipeline analytics.
\citet{xu2014path} presented an MTA model based on mutually exciting point process which independently considered the impressions and clicks as random events along the continuous time.
These exciting point process methods only considered the occurrence of the event which ignored the data of non-conversion cases.
% additive model
For analyzing the cumulative effects of the touch points, many works \cite{ji2017additional,zhang2014multi,xu2016lift,sinha2014estimating} made an assumption that the final conversion was influenced by the additive contributions from the touch points along the user browsing history.
However, it might result in a trend that more ad exposures were better which severely destroyed the user experience \cite{yuan2015supply}.
The reason is that the attribution of each touch points along the user behavior sequence may positively contribute or counteract the final conversion.
Thus, it is more reasonable to dynamically calculate and assign the attribution credits over the user behaviors.

% based on survival theory
Another school of MTA modeling is based on the survival theory \cite{zhang2014multi,ji2016probabilistic,ji2017additional}, which models the conversion event as the predictive goal and estimates the probability for the event occurrence at the specific time while considering the censored data, i.e., the true occurrence time is later than the observation time.
Nevertheless, these methodologies focus more on single point prediction and fail to consider the sequential patterns embedded in the user browsing history.
Moreover, the obtained attribution credits are mainly calculated based on heuristic additive assumptions, which may not be effective in practice.
They also made assumptions about the survival function such as exponential hazard function \cite{zhang2014multi,ji2017additional} and Weibull distribution for hazard rate estimation \cite{ji2016probabilistic} to make their model parameterized and thus optimizable. However, such parameterization could severely constrain the capacity of the model to fit various real-world data.

% deep learning and sequential prediction
Considering all the limitations above, we propose a dual-attention recurrent neural network for both conversion estimation and attribution. 
Attention mechanism is originally proposed for machine translation tasks, where a sequence-to-sequence model samples the next output word by attend each word of the input sentence \cite{bahdanau2014neural}.
In our problem, the attention is modeled as the attribution which may dynamically \textit{learn} to assign the credits over all the historical touch points for a specific user.
The sequential patterns have been efficiently captured by the recurrent mechanism.

Moreover, few works have discussed the budget allocation from the obtained attribution model. 
\citet{DiemertMeynet2017} proposed a bidding strategy based on the attribution credits for each real-time auction, which is not appropriate in general applications of online advertising.
\citet{geyik2014multi} presented a method for online budget allocation based on the obtained ROI from conversion attribution.
We borrow the idea of the ROI calculation from \cite{geyik2014multi} and devise an offline evaluation framework for multi-touch conversion attribution, which is the first offline experimental evaluation methodology for attribution models.

%\weinan{I suggest to write some sentences about attention network in deep learning.} \weinan{I add it by myself.}
\section{Methodology}\label{sec:methodology}
In this section, we firstly formulate the problem of the multi-touch conversion attribution, and then propose our sequential behavior modeling with dual-attention mechanism. Finally we present our evaluation protocol for conversion attribution guided budget allocation.

\subsection{Problem Definition}\label{sec:problem}
% problem definition and formulation
Without loss of generality, let us focus our study on the advertiser side.
When a user $u_i$ is taking Internet activities, e.g., browsing online contents, querying search engines or playing on social media, etc., there would be many sequential interactions between this user and the ad contents of an advertiser, which are called touch points for the ad campaign.
The observations are the user browsing sequences $\{ u_i, \{ q_j \}_{j=1}^{m_i}, y_i, T_i \}_{i=1}^{n}$ for each user $u_i$ who generates totally $m_i$ browsing activities with the ad of the advertiser.
$y_i$ is the indicator of whether the user converts and $T_i$ is the conversion time if the conversion occurs, otherwise null.
Each touch point $q_j = (\bx_j, z_j)$ contains the feature vector $\bx_j$ of the this touch point and the binary action type $z_j$, i.e., non-click impression or click.
Among them, the feature $\bx_j$ includes the side information of the user and the ad contents, e.g., user ID, advertising form, website, the operation systems and browser information, also with the channel ID feature $c_j$ over which this touch point is delivered and the time $t_j$ of the interaction occurrence.

The goal is to model the sequential user patterns and derive efficient conversion attribution credits for all the touch points $\{q_j\}_{j=1}^{m_i}$ along the user browsing sequence.
In return, the better conversion attribution obtained, the higher accuracy of the user conversion estimation for each browsing sequence. Similar formulation has been adopted in many literatures \cite{zhang2014multi,ji2016probabilistic,ji2017additional}.
In Sec.~\ref{sec:seq-model}, we present a recurrent neural network to model the sequential patterns and the final conversion rate.
We also apply sequence-to-sequence modeling for user click pattern mining and jointly learn impression and click patterns for conversion estimation.
The key component in this sequence modeling methodology is the dual-attention mechanism which takes two types of the user actions (i.e., impressions and clicks) into a unified comprehensive framework and facilitate the conversion modeling, which is described in Sec.~\ref{sec:attention}.
As a result, the obtained attention from the sequence modeling is naturally the conversion attribution over the whole user browsing history.

Interestingly, the derived conversion attribution also contributes to budget allocation for the subsequent ad delivery \cite{geyik2014multi}.
In Sec.~\ref{sec:eval-alg}, we propose an evaluation protocol for budget allocation with offline campaign data.

\subsection{Sequential Modeling}\label{sec:seq-model}
\begin{figure}[t]
  \centering
  \includegraphics[width=1.0\columnwidth]{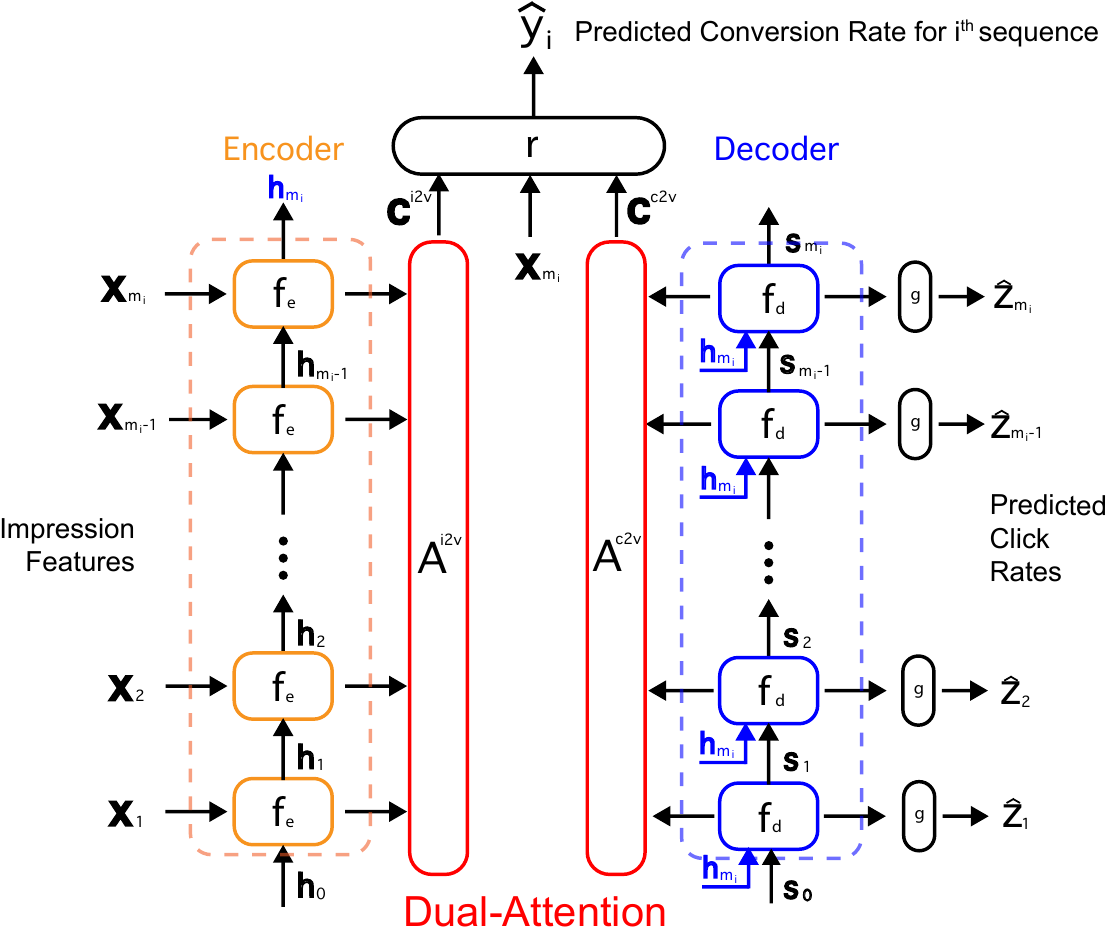}
  \caption{Sequential modeling with dual-attention.}\label{fig:model}
\end{figure}

% sequential prediction based method
We utilize recurrent neural network (RNN) for sequential user modeling, as illustrated in Figure~\ref{fig:model}.
Leveraging RNN for sequential modeling and time-series prediction has been widely applied in information retrieval systems \cite{song2017hierarchical,qin2017dual,zhai2016deepintent}.
Note that our methodology aims at final conversion estimation rather than sequential prediction for click at each touch point.

The whole structure can be divided as three separate parts that (i) the encoder for the impression-level behavior modeling; (ii) the decoder and sequential prediction for click probability; (iii) taking the above modeling output we implement dual-attention for jointly modeling impression and click behavior and produce the final conversion estimation.
We will clarify the first two parts in this section and discuss the attention mechanism later.

\minisection{Impression-level Behavior Modeling}
For the $i^{\text{th}}$ user behavior sequence $\{ u_i, \{ q_j \}_{j=1}^{m_i}, y_i, T_i \}$ where $q_j = (\bx_j, z_j)$, the input feature sequence to the RNN model is $\mathbf{x} = (\bx_1, \dots, \bx_j, \ldots, \bx_{m_i})$.
Since the side-information feature vector is mostly categorical \cite{zhang2016deep}, we firstly utilize an embedding layer to transform the sparse input feature into dense representation vector for subsequent training, which has been widely used in the related literatures \cite{qu2016product,wang2017dynamic}.

Then we feed the embedded feature vectors through the encoder RNN function $f_e$ approach as
\begin{equation}\label{eq:encoder}
\begin{aligned}
	\bh_j &= f_e(\bx_j, \bh_{j-1}) ~, \\
\end{aligned}
\end{equation}
where $\bh$ is the hidden vector of each time step $j$.
We implement $f_e$ as a standard long short-term memory (LSTM) model described in \cite{hochreiter1997long}, which has been widely used in natural language processing fields.

\minisection{Click-level Sequential Prediction}
In this part, our goal is to model the click action at each time when each ad is shown to the user.

In the sequence-to-sequence model, the decoder defines a probability over the click outcomes $\mathbf{z}$ by decomposing the \textit{joint probability} into the ordered conditionals as
\begin{equation}
	p(\mathbf{z}) = \prod_{j=1}^{m_i} p( z_j=1 ~|~ \{ z_1, \ldots, z_{j-1} \}, \mathbf{x}) ~.
\end{equation}
Note that $\mathbf{z} = (z_1, \ldots, z_{m_i})$ and $\mathbf{x} = (\bx_1, \ldots, \bx_{m_i})$.
With this decoder each conditional probability is modeled as
\begin{equation}
	\hat{z}_j = p(z_j=1 ~|~ \{z_1, \ldots, z_{j-1} \}, \mathbf{x}) = g(z_{j-1}, \bss_j) ~.
\end{equation}
Here $g$ is the output function which is a multi-layer fully connected perceptron with sigmoid activation function $\text{sigmoid}(x) = \frac{1}{1+e^{-x}}$ that outputs the probability of $z_j=1$.
And $\bss_j$ is the hidden vector at click-level of the $j^{\text{th}}$ touch point, calculated by
\begin{equation}\label{eq:decoder}
	\bss_j = f_d(\bss_{j-1}, z_{j-1}, \bh_{m_i})~,
\end{equation}
where $f_d$ is the nonlinear decoder RNN function, potentially multi-layered, that models the sequential click patterns for user behavior sequence. We utilize the same structure of LSTM model as the encoder $f_e$.
Each hidden state $\bss_j$ in the decoder uses the last hidden state $\bh_{m_i}$ from the encoder.
Our first loss is based on the sequential prediction for click probabilities as
\begin{equation}\label{eq:click-obj-func}
L^c = \sum_{i=1}^n \sum_{j=1}^{m_i} -z_j \log \hat{z}_j - (1-z_j) \log (1-\hat{z}_j) ~.
\end{equation}

There are two rationales for the sequence-to-sequence click prediction in this work.
The first is to some extent similar with the idea of multi-task learning to alleviate the data sparsity problem and conduct a shared base representation of user behavior features.
As is known that the users follow a pattern of actions that they may click after impression of the ad and after a sequence of ad delivery they may (not) drive the final conversion, which derives the data sparsity problem behind the ``impression-click-conversion'' action pattern \cite{ma2018entire}.
Specifically, clicks are less frequent events than impressions and conversions are much rarer than clicks.
It is necessary to conduct a methodology to tackle with the data sparsity challenge.
Our intuition is to utilize the signal of click behavior to boost the estimation capacity for the sparse conversion behaviors.
Another reason for the sequential click pattern mining is to obtain the click-level attribution modeling for multiple pre-conversion behavior modeling, which has shown statistically more important attribution credits than impression-level behaviors in our experimental results.

\subsection{Learning Attribution with Dual-attention}\label{sec:attention}
% attention as attribution
Our final goal is to model the sequential user patterns and predict the conversion probability.
The final output is calculated as
\begin{equation}\label{eq:final-conv-estimation}
\begin{aligned}
	\bc^{i2v} &= A^{i2v}({\bh_1, \ldots, \bh_j, \ldots, \bh_{m_i}}) ~,\\
	\bc^{c2v} &= A^{c2v}({\bss_1, \ldots, \bss_j, \ldots, \bss_{m_i}}) ~, \\
	\hat{y}_i &= p(y=1 ~|~ \mathbf{x}, \mathbf{z}) = r(\bx_{m_i}, \bc^{i2v}, \bc^{c2v}) ~,
\end{aligned}
\end{equation}
where $r$ contains a weighting function for balancing impression-level and click-level attribution, which will be described in detail later in this section, and a dense multi-layer neural network for the final conversion prediction.
$\bx_{m_i}$ is the feature vector of the last touch point which would be fed through the same embedding layer as that in Sec.~\ref{sec:seq-model}. $\bc^{i2v}$ is the context vectors of all the input user behavior vectors capturing impression patterns and $\bc^{c2v}$ is the context vector from modeling the click patterns for conversion estimation.

\minisection{Learning Attention through Conversion Estimation} The loss is calculated by the cross entropy for the conversion estimation that
\begin{equation}\label{eq:obj-func}
	L^v = \sum_{i=1}^n - y_i \log \hat{y}_i  - (1-y_i) \log(1 - \hat{y}_i) ~.
\end{equation}

\begin{figure}[t]
  \centering
  \includegraphics[width=0.65\columnwidth]{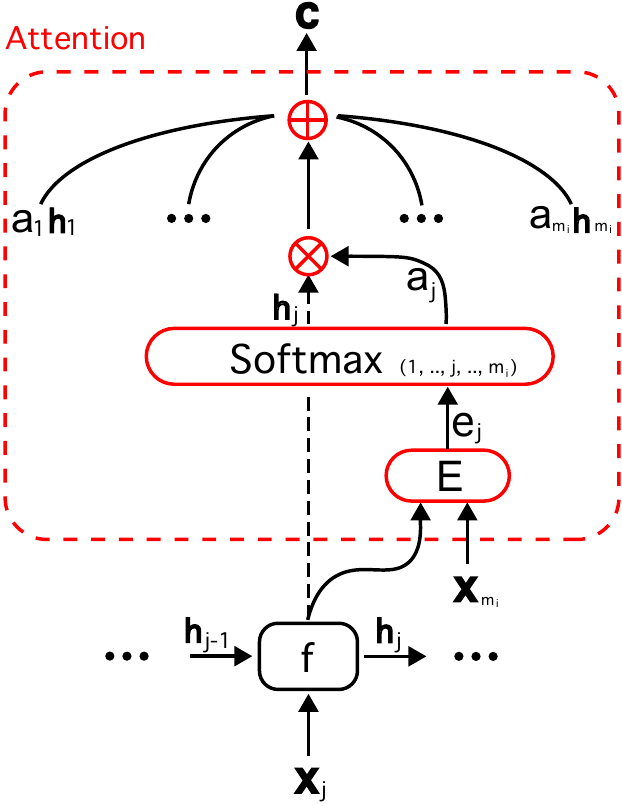}
  \caption{Attention calculation mechanism.}\label{fig:attention}
\end{figure}

The key component is the attention input $\bc^{i2v}$ and $\bc^{c2v}$ from impression-level and click-level, respectively.
The mechanism of attention function $A$ is illustrated in Figure~\ref{fig:attention}.

To calculate the impression-to-conversion attention $\bc^{i2v}$ and the click-to-conversion attention $\bc^{c2v}$, we propose a unified energy-based function as
\begin{equation}
\begin{aligned}
	\bc &= l({\bh_1, \ldots, \bh_j, \ldots, \bh_{m_i}}) = \sum_{j=1}^{m_i} a_j \bh_j ~.
\end{aligned}
\end{equation}
Note that this formulation is expressed to calculate $\bc^{i2v}$ while without losing generality by replacing $\bx$ and $\bh$ in Eq.~(\ref{eq:encoder}) with $\bss$, $z$ and $\bh_{m_i}$ in Eq.~(\ref{eq:decoder}) for $\bc^{c2v}$ calculation.

And the weight $a_j$ is calculated based on the softmax operated energy value $e_j$ as
\begin{equation}\label{eq:attn-weight}
	a_j = \frac{\exp(e_j)}{\sum_{k=1}^{m_i} \exp(e_k)} ~,
\end{equation}
where
\begin{equation}\label{eq:attn-energy}
	e_j = E(\bh_j, \bx_{m_i})
\end{equation}
is an energy model which scores the credit of each touch point to the final conversion.
Note that the energy function $E$ is a nonlinear multi-layer deep neural network with tanh activation function $\text{tanh}(x)=\frac{e^x - e^{-x}}{e^x + e^{-x}}$.
The way we calculate the attention through the energy function $E$ is similar to that in the natural language processing field \cite{bahdanau2014neural,gehring2017convolutional}.

As a result, the dual-attention mechanism is expressed as
\begin{equation}
\begin{aligned}
	\bc^{i2v} &= \sum_{j=1}^{m_i} a^{i2v}_j \bh_j ~, \\
	\bc^{c2v} &= \sum_{j=1}^{m_i} a^{c2v}_j \bss_j ~, \\
\end{aligned}
\end{equation}
and the values of $a$ in both attention calculation are obtained through Eqs.~(\ref{eq:attn-weight}) and (\ref{eq:attn-energy}).

\minisection{Attribution Calculation with Dual-attention} Till now, we have obtained the estimated conversion probability $p(y|\mathbf{x}, \mathbf{z})$ and the attention results $a^{i2v}_j$ and $a^{c2v}_j$ for each touch point, based on which we can naturally assign the credits for each touch point $j$.

Recall that the final conversion estimation is based on the learned dual-attention vector, i.e., $\bc^{i2v}_j$ and $\bc^{c2v}_j$, and the final touch point feature vector $\bx_{m_i}$, here we adopt a dynamic weighting function $f_{\lambda}$ to balance the effcts of the two attentions that
\begin{equation}\label{eq:lambda-calculation}
	\lambda = \frac{\exp\left[f_{\lambda}(\bx_{m_i}, \bc^{c2v})\right]}{\exp\left[f_{\lambda}(\bx_{m_i}, \bc^{i2v})\right] + \exp\left[f_{\lambda}(\bx_{m_i}, \bc^{c2v})\right]} ~,
\end{equation}
where $\lambda$ measures the importance of the click-level attention w.r.t. that from the impression-level and $f_{\lambda}$ is a multi-layer perceptron whose goal is to learn the weight of two attention results for the final conversion estimation.

Thus, the estimation function $r$ mentioned first in Eq.~(\ref{eq:final-conv-estimation}) is that
\begin{equation}
	r(\bx_{m_i}, \bc^{i2v}, \bc^{c2v}) = r_{\text{conv}}\left( (1 - \lambda) \cdot \bc^{i2v} + \lambda \cdot \bc^{c2v} \right) ~.
\end{equation}
Here $r_{\text{conv}}(\cdot)$ is a multi-layer neural network for conversion estimation with sigmoid activation function.

Now that we have weighted the contribution of impression-level and click-level attentions for final conversion estimation, we can naturally obtain the attribution for each touch point through these learned patterns as 
\begin{equation}\label{eq:attention-calc}
	\text{Attr}_j = (1-\lambda) \cdot a^{i2v}_j + \lambda \cdot a^{c2v}_j ~.
\end{equation}

The motivation for building such a dual-attention mechanism is that we care both the impression-level and the click-level user behavior patterns to facilitate conversion estimation and the subsequent attribution results.

\subsection{Evaluation Protocol}\label{sec:eval-alg}
% attribution as direction 
% replay method
With the attribution credits allocated to the touch points along the user behavior sequences, our focus moves onto the efficiency of budget allocation based on the calculated attribution credits.
Note that almost all the related works report only the conversion estimation performance; few of them test the budget allocation under the obtained attribution credits, except online A/B testing which is expensive and risky.
Here we propose a framework to offline evaluate the conversion attribution model based on the historic data of a campaign.

In online advertising, the guideline of the advertiser to allocate budgets for the subsequent ad delivery on different channels of the ad campaign is intuitively based on the past performance.
The performance here means the effectiveness, i.e., return on investment (ROI), of the ad delivery onto each channel, which is measured as the obtained positive user conversions w.r.t. the delivered ad costs.
The most intuitive idea is to allocate more budgets for the channels or sub-campaigns with higher ROI than others, to gain more user conversions.
However, different attribution methods substantially influence the ROI calculation results \cite{geyik2014multi}.
Specifically, the idea of our evaluation protocol is to first calculate the ROI performance results for each channel under different attribution models, and then utilize the offline replay of ad delivery history to measure the performance of the obtained fresh conversions and, considering the costs in the offline replay, calculate the effectiveness results of the ad delievery for different evaluation baselines.
So that, under this evaluation, the more proper attribution credits one model proposed, the better performance it would obtain in the subsequent budget allocation for different ad channels and naturally obtain better performance through the ad replay evaluation.

Next we will first present the budget allocation method based on attribution-guided ROI results.
Then we illustate our back evaluation algorithm w.r.t. the allocated budget scheme for later performance comparison.

\minisection{ROI-based Budget Allocation} In this stage, the first problem is to allocate the budget $\{ b_1, \ldots, b_K \}$ across $K$ channels according to the obtained attribution credits.
Here we follow the idea presented in \cite{geyik2014multi} that
\begin{equation}\label{eq:roi-calc}
	\text{ROI}_{c_k} = \frac{\sum_{\forall y_i=1} \text{Attr}(c_k | y_i) ~ V(y_i)}{\text{Money spent on channel} ~ c_k} ~,
\end{equation}
where 
\begin{align}
\text{Attr}(c_k | y_i) = \sum_{j=1}^{m_i} \text{Attr}_j \cdot 1(c_j = c_k)
\end{align}
is the overall credit attributed on the channel $c_k$ by aggregating the credit $\text{Attr}_j$ of all touch points $j$'s within this channel, $1(\cdot)$ is the indicator function, and $V(y_i)$ is the value of the conversion.
After the ROI calculation we allocate budgets for different channels w.r.t. the obtained ROI proportion as $b_k = \frac{\text{ROI}_{c_k}}{\sum_{v=1}^K ROI_{c_v}} \times B$ for channel $c_k$, where $B$ is the total budget.

\minisection{Back Evaluation under Reallocated Budgets} The historic data is a series of event sequences $\{\text{seq}^s\}$ and each sequence is represented as $\{ (q_i, \bx_i, t_i, y_i, c_i, o_i) \}$ where each user interaction identified by $q_i$ is on the specific channel $c_{i}$ at time $t_i$ with cost $o_i$ and the feature vector $\bx_i$ includes the click label information.
Each event is either an ad serving event without conversion ($y_i = 0$) or a user conversion event ($y_i = 1$).

In addition, we introduce the concept of \emph{conversion blacklist}.
If the budget of one channel $c_k$ is exhausted at moment $t$, then the conversion events of all the unfinished sequence with ad serving event after $t$ on channel $c_k$ become invalid.
These conversions should be put into the conversion blacklist. This is reasonable because if the user cannot observe the ad touch point, it is no guarantee that she will finally convert at the end of the sequence. Such a back test result serves as a lower bound estimation of the true but unknown performance.

Given the budget allocation $(b_1, b_2, \ldots, b_K)$ across $K$ channels, we can make the following back test as presented in Algorithm \ref{alg:eval}. 
Specifically, the back test goes over the historic events by their recording time $t_i$. 
If there is no budget left for the channel $c_k$ for the back playing event  $(q_i, \bx_i, t_i, y_i, c_i=c_k, o_i) $, then put the sequence indicator $\text{seq}^s$ into conversion blacklist.
After the back test, we can evaluate the attribution models by the cost $O$ and the obtained valid conversion number $Y$.

\begin{algorithm}[t]
	\caption{Back Evaluation for Budget Allocation}\label{alg:eval}
	\begin{algorithmic}[1]
		\REQUIRE
		The events $\{ (q_i, \bx_i, t_i, y_i, c_i, o_i) \}$ ordered by the\\
		serving time $t_i$ and the budget allocation $\{ b_1, \ldots, b_K \}$.
		\ENSURE
		The total conversion number $Y$ and the total cost $O$.
		\STATE Initially set the blacklist of sequence list $\mathcal{B} = \{\}$ and the obtained conversion number $Y=0$, total cost $O=0$.
		\FOR{each event $(q_i, \bx_i, t_i, y_i, c_i, o_i)$ in the data}
		\IF{$\text{seq}^s$ not in $\mathcal{B}$}
			\IF{the budget for $c_i$ channel $b_{c_i} > o_i$}
			\STATE $O = O + o_i$
			\STATE $Y = Y + y_i$
			\STATE $b_{c_i} = b_{c_i} - o_i$
			\ELSE
			\STATE Put $\text{seq}^s$ into $\mathcal{B}$
			\ENDIF
		\ENDIF
		\ENDFOR
	\end{algorithmic}
\end{algorithm}

\section{Experiments}\label{sec:exp}
% experiment introduction
% two-stage experiment
In this section, we firstly present the experiment setup including the description of two real-world datasets, the evaluation measurements and the compared models used in our experiments.
Then we illustrate the corresponding results for the two-staged experiment settings.
The first stage is for the conversion estimation accuracy while the second one is for the attribution guided budget allocation performance over history data.
In addition, we have published our code\footnote{Repeatable experiment code link: \href{https://github.com/rk2900/deep-conv-attr}{https://github.com/rk2900/deep-conv-attr}.} for repeatable experiments.

\subsection{Datasets}
% datasets
% miaozhen & criteo
% please cite the criteo paper when describing the criteo dataset.
In our experiments, we apply our model and the compared baselines over two real-world datasets.

\vspace{5pt} \noindent \textbf{Miaozhen} is a leading marketing technique company in China. This dataset \cite{zhang2014multi} includes almost 1.24 billion advertising logs from May 1 to June 30 and April 4 to June 9 in 2013. Specifically it contains about 59 million users and 1044 conversions. These ad contents have been exposed over 2498 channels with 40 advertising forms, such as button ads and social ads.
In the dataset, every time a user is exposed to the ad or click on the ad contents, the exact time of the user action with the side information will be recorded.
Moreover, it also contains the purchasing information as the conversion of the user with the corresponding timestamp.
The user is tracked according to the user cookie identifier which is anonymized in the dataset.
With these logs, we are able to reconstruct the time line of the user action sequence including impression and click information, the exposure ad channels and the conversion labels for each sequence.

\vspace{5pt} \noindent \textbf{Criteo} is a pioneering company in online advertising research. They have published this dataset\footnote{Processed dataset link: \href{http://apex.sjtu.edu.cn/datasets/13}{http://apex.sjtu.edu.cn/datasets/13}.}
for attribution modeling in real-time auction based advertising \cite{DiemertMeynet2017}.
This dataset is formed of Criteo live traffic data in a period of 30 days.
It has more than 16 million impressions and 45 thousand conversions over 700 campaigns.
The impressions in this dataset may derive click actions so each touch point along the user action sequence has a label of whether a click has occurred, and the corresponding conversion ID if this sequence of touch points leads to a conversion event.
Each impression log also contains the cost information, which will be used in our second state experiment for attribution effectiveness evaluation.
Since the channel data are missing so we take campaign as the budget allocation targets.

\minisection{Data Preprocessing and Sampling} Since the user conversion is a rare event, we perform negative sampling in data preprocessing.
Following \cite{zhang2014multi,ji2017additional}, the sequence preparation and sampling rules are that (i) if the user has multiple conversion events, her action sequence will be split according to the conversion time to guarantee that each sequence has at most one conversion; (ii) we extract the user action sequences with the minimal length of 3 and maximal length of 20 with the sequence duration within 14 days; (iii) all of the user sequences leading to conversion events have been retained and we uniformly sample the sequences without conversions to 20 times of the number of converted sequences.

\subsection{Evaluation Pipeline and Metrics}
Here we present the evaluation pipeline and the measurements over the compared settings. Overall, we have two stages of the experiments.

The first stage focuses on the conversion estimation performance which has been widely adopted in the conversion attribution task \cite{zhang2014multi,ji2016probabilistic,ji2017additional}.
Specifically, given the evaluation samples $\{ u_i, \{ q_j \}_{j=1}^{m_i}, y_i, T_i \}_{i=1}^{n}$ in the test dataset, the model predicts the output of conversion probability $\hat{y}_i$ after the user going through each sequence of touch points.
There are two evaluation metrics for measuring the performance of each model.
\textbf{Log-loss} is the common measurement to estimate the classification performance for the event probability prediction which is the cross entropy as is expressed in Eq.~(\ref{eq:obj-func}).
The other metric is \textbf{AUC} (area under ROC curve) which measures the pairwise ranking performance of the classification results between the converted and nonconverted sequence samples.

The second stage aims at the performance of budget allocation, with the calculated conversion attributions, for various channels or sub-campaigns.
According to Algorithm~\ref{alg:eval}, we replay all the test campaign data w.r.t. the recorded timestamp, and calculate the performance for the below metrics.
Note that we set the evaluation budgets for each model as 1/2, 1/4, 1/8, 1/16, 1/32 of the total budget in the whole test dataset.
The similar evaluation setting has been widely adopted in online advertising researches \cite{zhang2014optimal,ren2016user,ren2017bidding}.
The \textbf{number of conversions} is the total number of the achieved conversions. \textbf{Profit} is the total gains, i.e., total value of the obtained conversions.
The other two metrics are \textbf{CVR} (conversion rate) and \textbf{CPA} (cost per conversion action).
CVR is the ratio of the converted sequences among all the touched user impression sequences which reflects the ratio of gain for the ad delivery.
And CPA is the cost averaged by the obtained conversion numbers which mesures the efficiency for the ad campaign.
Note that only Criteo dataset contains the cost information so that our second stage experiments is conducted on Criteo dataset.

\subsection{Compared Settings}
% baselines
% LR (Xuhui Shao), AH, AMTA, dual-attention recurrent neural network (DARNN)
In this section, we discuss the compared baselines and our model settings.
We compare four baseline models with our dual-attention model.
We also discuss the advantages of our dual-attention mechanism against the normal RNN model with single attention mechanism.
Note that the click label ground truth $z$ has been included in the input feature $\bx$ in the other baseline models, except for our proposed model which utilizes click as the prediction label, for equally comparison.
% \yuchen{We included the click label when training our model on Criteo.}.

\begin{itemize}[leftmargin=0.8mm]
\item \textbf{LR} is the Logistic Regression model proposed in \cite{shao2011data} and the attribution is calculated as the learned coefficient values of the regression model parameter for each channel.
\item \textbf{SP} is a Simple Probabilistic model whose idea is derived from \cite{dalessandro2012causally} and the conversion rate of each user action sequence is calculated as in \cite{zhang2014multi} that
\vspace{-5pt}
\begin{equation}
	p( y = 1 | \{ c_j \}_{j=1}^{m_i} ) = 1 - \prod_j^{m_i} ( 1 - \text{Pr}(y = 1 | c_j = k) ) ~,
\vspace{-5pt}
\end{equation}
where $\text{Pr}(y = 1 | c_j = k)$ is the conversion probability from the observed data w.r.t. the $k\text{th}$ channel.
\item \textbf{AH} (AdditiveHazard) model is the first work \cite{zhang2014multi} using survival analysis and additive hazard function of conversion with the consideration of the touch point time to predict the final conversion rate. More details could be found in the paper.
\item \textbf{AMTA} is the Additional Multi-touch Attribution model proposed in \cite{ji2017additional} which was state-of-the-art for this conversion attribution problem. It applies survival analysis to model the conversion estimation and utilizes the hazard rate of conversion at the specific time to model the conversion attribution.
\item \textbf{ARNN} is the normal Recurrent Neural Network (i.e., only encoder part) method with the single Attention merely based on impression-level patterns to model the conversion attribution that $\hat{y}_i = r'(\bc^{i2v})$, rather than sequence-to-sequence modeling in Eq.~(\ref{eq:decoder}). This model is to illustrate the advantage of our dual-attention mechanism for data sparsity problem and multi-view learning schema.

\item \textbf{DARNN} is our proposed model with dual-attention mechanism, which has been described in Section~\ref{sec:methodology}.
\end{itemize}
All the deep models are trained separately over one NVIDIA GeForce GTX 1080 Ti with Intel Core i7 processor for five hours.
The detailed hyperparameter settings have been described in our published code, including learning rate, feature embedding size, hidden state size in RNN cell, etc.

\begin{table}[h]
	\centering
	\caption{Conversion estimation results on two datasets. AUC: the higher the better; Log-loss: the lower the better.}\label{tab:conv_est}
	\vspace{-5pt}
	\resizebox{0.65\columnwidth}{!}{
		\begin{tabular}{c|rr|rr}
			\hline
			& \multicolumn{2}{c|}{Miaozhen} & \multicolumn{2}{c}{Criteo} \\
			Models & AUC & Log-loss & AUC & Log-loss \\
			\hline
			LR & 0.8418 & 0.3496  & 0.9286 & 0.3981 \\
			SP&0.7739&0.5617&0.6718&0.5535\\
			AH & 0.8693 & 0.6791 & 0.6791 & 0.5067  \\
			AMTA & 0.8357 & 0.1636  & 0.8465  & 0.3897   \\
			ARNN & 0.8914 & 0.1610 & 0.9793 & 0.1850  \\
			DARNN & \textbf{0.9123} & \textbf{0.1095} & \textbf{0.9799} & \textbf{0.1591}  \\
			\hline
	\end{tabular}}
	\vspace{-10pt}
\end{table}

\subsection{Conversion Estimation Performance}
% convertion estimation performance
% list the results of the above models over these metrics

Our first evaluation is to measure the performance of user conversion estimation.
Table~\ref{tab:conv_est} presents the detailed evaluation results under different models.
From the statistics in the table, our model outperforms other baselines under both evaluation metrics.
The results also reflect the other findings as below.
(i) Both of the attention-based methods, i.e., DARNN and ARNN, achieve much better performance for sequential prediction than other compared models, which reflects the great pattern mining capability of deep neural networks.
(ii) The exciting point process based methods AH and AMTA has poor classification performance for the conversion estimation. The reason is that they are designed to model the additive hazard ratio of conversion for each touch point. Though they learn the conversion prediction for the whole sequence, they do not consider much of the sequential patterns within the user behavior sequence.
(iii) For the log-loss metric, the baselines get relatively higher (i.e., poorer) values than the deep models, which reflects that these baselines predict the conversion probability with totally large or small absolute values. Note that, however, AUC has no relationship to the direct output value of the model but considers the pairwise ranking performance. So almost all the baselines get considerably acceptable AUC results.

% learning curves
\begin{figure}[h]
	\centering
	\includegraphics[width=1\columnwidth]{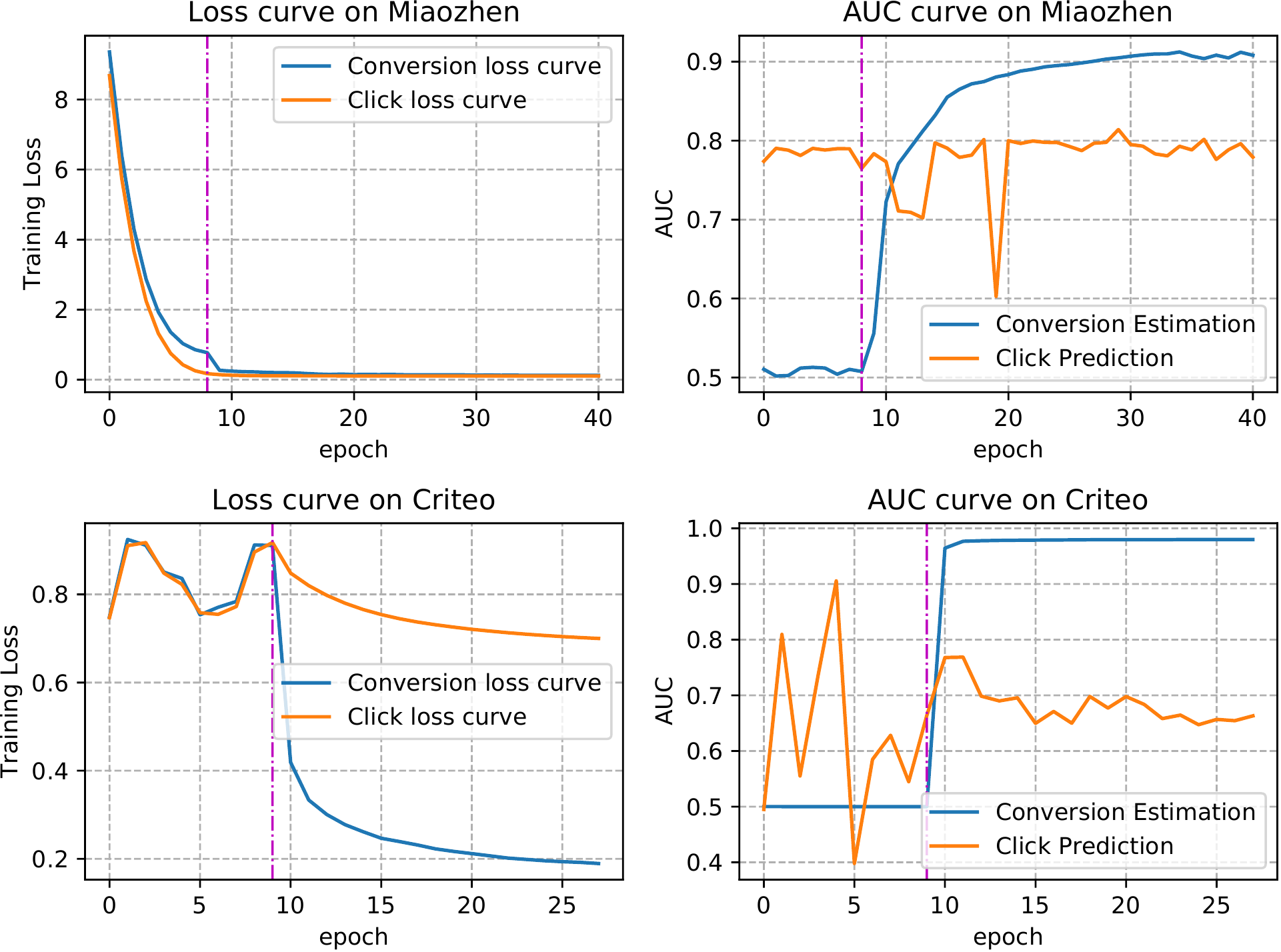}
	\caption{Learning curves on two datasets. Here one ``epoch'' means one whole iteration over the train dataset. The vertical purple line means the conversion estimation optimization starts in the second training stage.}\label{fig:learning_curve}
	\vspace{-5pt}
\end{figure}

As for the learning procedure, since our proposed DARNN model captures both impression-level and click-level patterns and optimizes under two types of losses as that in Eqs.~(\ref{eq:click-obj-func}) and (\ref{eq:obj-func}), the training procedure of our DARNN model generates two learning curves in Figure~\ref{fig:learning_curve}.
In model training, we firstly make the model learn the click patterns, i.e., only optimize under the sequential click prediction loss as that in Eq.~(\ref{eq:click-obj-func}) and then, after the convergence of the first objective, we turn on the conversion estimation training, i.e., optimize under both two losses till convergence of the conversion loss.
The convergence of each objective is defined as two successive rising of the optimization loss.
The reason of two-stage training is to stabilize the model optimization under these two learning objectives.
The similar training procedure has been studied in multi-task learning for recommender systems \cite{cao2018neural}.

We may easily find from Figure~\ref{fig:learning_curve} that our model not only optimizes the sequential click prediction, but also learns the conversion estimation.
Moreover, the two learning objectives have been alternatively optimized to convergence at the second stage.
Both the click prediction and the conversion estimation achieve excellent prediction performance.

\begin{table*}[h]
	\centering
	\caption{Budget allocation evaluation results. CPA: the lower the better; Profit \& CVR: the higher the better.}\label{tab:budget_allo}
	%\scriptsize
	\resizebox{\textwidth}{!}{
		\begin{tabular}{c|ccccc|ccccc|ccccc|ccccc}
			\hline
			& \multicolumn{5}{c|}{CPA}  & \multicolumn{5}{c|}{Profit} & \multicolumn{5}{c|}{Conversion Num.} & \multicolumn{5}{c}{CVR}\\
			Models & $\frac{1}{2}$ & $\frac{1}{4}$& $\frac{1}{8}$ &$\frac{1}{16}$ & $\frac{1}{32}$ & $\frac{1}{2}$ & $\frac{1}{4}$& $\frac{1}{8}$ &$\frac{1}{16}$ & $\frac{1}{32}$& $\frac{1}{2}$ & $\frac{1}{4}$& $\frac{1}{8}$ &$\frac{1}{16}$ & $\frac{1}{32}$& $\frac{1}{2}$ & $\frac{1}{4}$& $\frac{1}{8}$ &$\frac{1}{16}$ & $\frac{1}{32}$\\
			\hline
			LR &31.79 &29.47& 29.77 & 27.83& 27.46 & 8.022&6.938& 4.386 & 3.238 & 1.954  & 576&427&275 & 181& 107  &0.0928&0.0910& 0.0873& 0.0827 & 0.0748  \\
			SP & 24.84 &22.98& 21.29& 21.39& 20.93 & 13.07&10.28&7.694&4.648&2.776&452& 315& 191& 112,&62&0.1205&0.1251&0.1223&0.1122&0.1028\\
			
			AH &24.69&21.84& 20.37 &18.89& 19.32 &27.03&22.08 & 15.38 & 10.32 & 5.491 &1286&925 & 607 & 385 & 208 & 0.1120&0.1194 & 0.1197 & 0.1183 & 0.1079   \\
			AMTA &24.71&21.91& 20.43 & 18.89 & 19.41 &27.01&21.96 & 15.29 & 10.33 & 5.446 &1285&922 & 605 & 385 & 207 &0.1118&0.1192 & 0.1195 &0.1183& 0.1073\\
			ARNN &26.66&23.98& 22.61 & 19.86 & 18.96 &29.10&\textbf{23.32} & 15.81 & 11.68 & \textbf{7.010} &\textbf{1527}&\textbf{1073} & \textbf{684} & \textbf{452} & \textbf{262}&0.1073&0.1137 & 0.1119 &0.1206& 0.1174 \\
				
			DARNN & \textbf{23.47 } & \textbf{21.24} & \textbf{18.50} & \textbf{16.85} & \textbf{17.63} & \textbf{29.25} & 22.56& \textbf{17.58} & \textbf{12.09} & 6.26 & 1315 & 922 & 646 & 419 & 223 & \textbf{0.1226} & \textbf{0.1274} & \textbf{0.1339} & \textbf{0.1321} & \textbf{0.1206}\\	
			\hline
	\end{tabular}}
\end{table*}

\subsection{Attribution Guided Budget Allocation}
\begin{figure}[t]
	\centering
	\includegraphics[width=1.0\columnwidth]{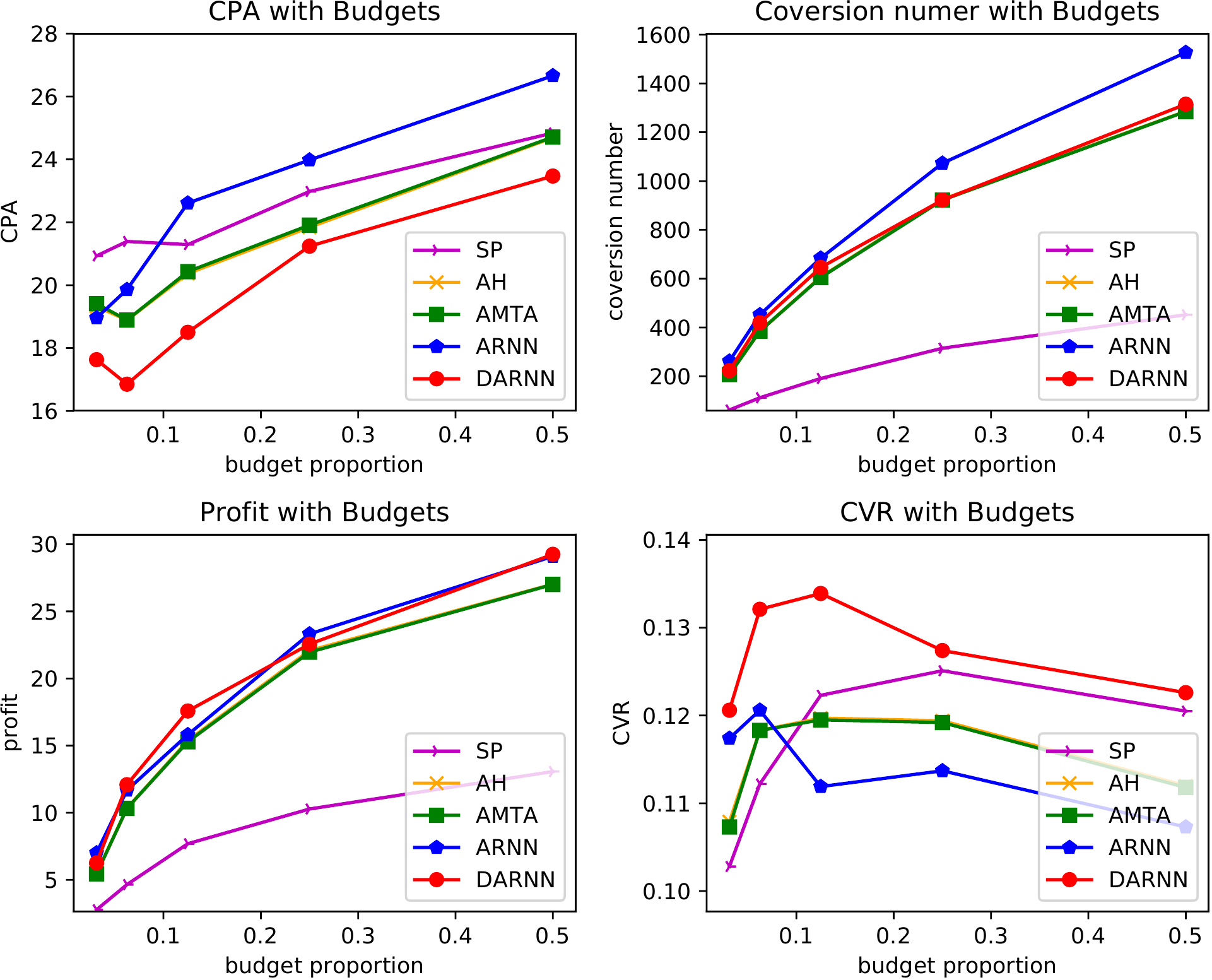}
	\caption{Performance with budgets on Criteo.}\label{fig:budget_allocation}
	\vspace{-10pt}
\end{figure}

% attribution directs budget allocation and replay performance
% the CPA with conversion number and cost in criteo dataset
In the second stage of the experiments, we evaluate the effectiveness of different conversion attribution models for budget allocation.
After replaying the historic touch points along the ordered timestamps, we calculated the total costs and the obtained conversion numbers of the compared model settings.
To calculate the obtained profits for each model, we make the conversion value $V(y_i=1)$ in Eq.~(\ref{eq:roi-calc}) as eCPA (effective cost-per-action) which is constant for each model and calculated as $V(y_i=1) = \text{eCPA}_{\text{train}} = (\text{total cost} / \text{conversion number})$ in the training data.
The detailed results are presented in Table~\ref{tab:budget_allo} and Figure~\ref{fig:budget_allocation}.
As is presented in the table, since LR performs quite poor, we eliminate LR results in the figure for better illustration.
Moreover, note that, we also compared simple \textit{last-touch} attribution method in the second-stage experiment.
We did not report this heuristic method in our experiments since the result showed that AH baseline model performed almost the same as the last-touch attribution method which is quite interesting and needs further investigations in the future work.

From Table~\ref{tab:budget_allo} and Figure~\ref{fig:budget_allocation}, we may find that:
(i) As the budget increases, all the models spend more to earn each user action, i.e., the CPA value of each model is increasing, which is reasonable.
(ii) Both of the attention-based neural network models, i.e., ARNN and DARNNs, achieve relatively better performance compared with the other models over all the evaluation metrics. The reason is probably the sequential pattern mining of these two models.
(iii) The two baselines AMTA and AH achieve very similar performance, which is probably accounted for the similar idea of the additional conversion probability modeling within their models.
(iv) DARNN model achieves the best performance under CPA and CVR, which reflects the effectiveness our learned attribution values of dual-attention mechanism. Moreover, this result also shows the advantage of the dual attention mechanism over single attention model ARNN.
(v) ARNN spends money more aggressive than other models thus getting poor CPA result. The reason may be that its attribution is based merely on impression-level and the pattern captured tends to long-term investment on the user behavior. However, our DARNN model spends the budget more economically which leads to more efficient budget pacing, i.e., lower CPA. This indicates that combining both impression-level and click-level attention will take advantages of both long-term (impression to conversion) and short-term (click to conversion) behavior patterns.

\subsection{Comprehensive Analysis}
% case study of the learned attribution (telling stories)
In this part, we look deeper into the learned attribution model.
We first discuss the calculated attribute credits over both touch point level and channel level, and then analyze the results of the learned weighting parameter $\lambda$ according to Eq.~(\ref{eq:lambda-calculation}) which controls the influence of the two types of user actions, i.e., impressions and clicks.

% hyperparameter tuning e.g., \lambda tuning.
First, we illustrate the touch point attribution in Miaozhen dataset which calculates the averaged attribution credits over all the sequence samples with fixed sequence length, on each touch point.
Specifically, here the credits on the $j^{\text{th}}$ touch point is averaged over all the converted sequences with the fixed sequence length $m$ as $\overline{\text{Attr}}_j = \frac{1}{N_m} \sum_{i=1}^{N_m} y_i \cdot \text{Attr}_{ij}$,
where $N_m$ is the total number of the sequences with length $m$ and $y_i$ is the conversion indicator of the sequence sample.

\begin{figure}[t]
	\centering
	\includegraphics[width=1.0\columnwidth]{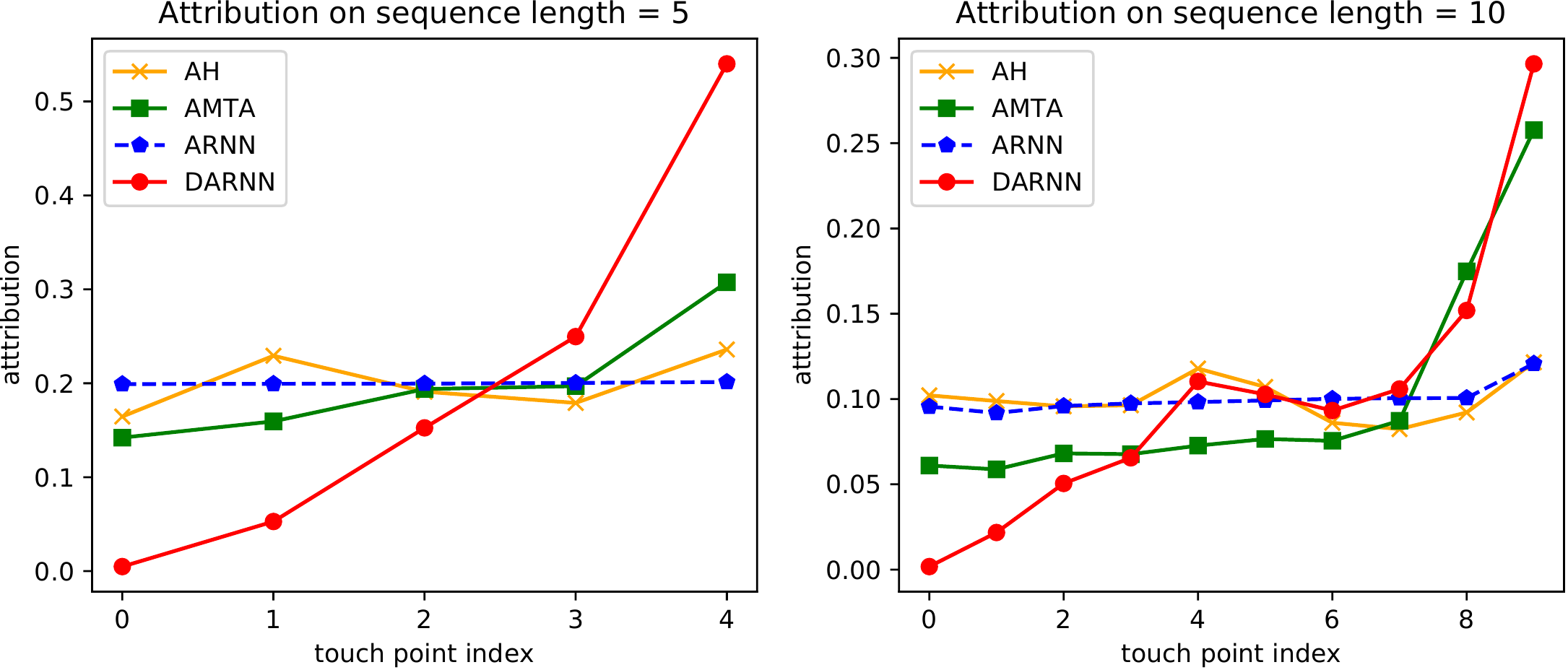}
	\caption{Touch point level attribution statistics (Miaozhen).}\label{fig:attribution_over_sequence}
	\vspace{-5pt}
\end{figure}
Figure~\ref{fig:attribution_over_sequence} illustrates the touch point conversion attribution results on the sequences with length of 5 and 10 respectively.
Since LR and SP calculate the attribution based on different channels rather than each touch point, so we cannot get the specific result of these two models at the touch point level.
From the figure we may find that the credits attributed on each touch point varies over different models.
When sequence length is relatively short, DARNN learns that the touch point closer to the final touch may more likely derive the final conversion.
In longer sequences (with length of 10), our DARNN model place higher credits for the touch points in the middle process while the attribution drops a little later and consequently rises to much higher when final conversion approaches.
This phenomenon is reasonable since it is not always correct about endless ad delivery for the user and, moreover, it reflects the tradeoff between the ad effectiveness and the user experience of the Internet service.
However, ARNN seems to ``average'' the credits over all the touch points within the sequence. Note that ARNN only concerns impression-level contributions, which in contrast shows the great effects of click-level patterns in our proposed dual-attention mechanism on the final conversion attribution.

\begin{figure}[t]
	\centering
	\includegraphics[width=1.0\columnwidth]{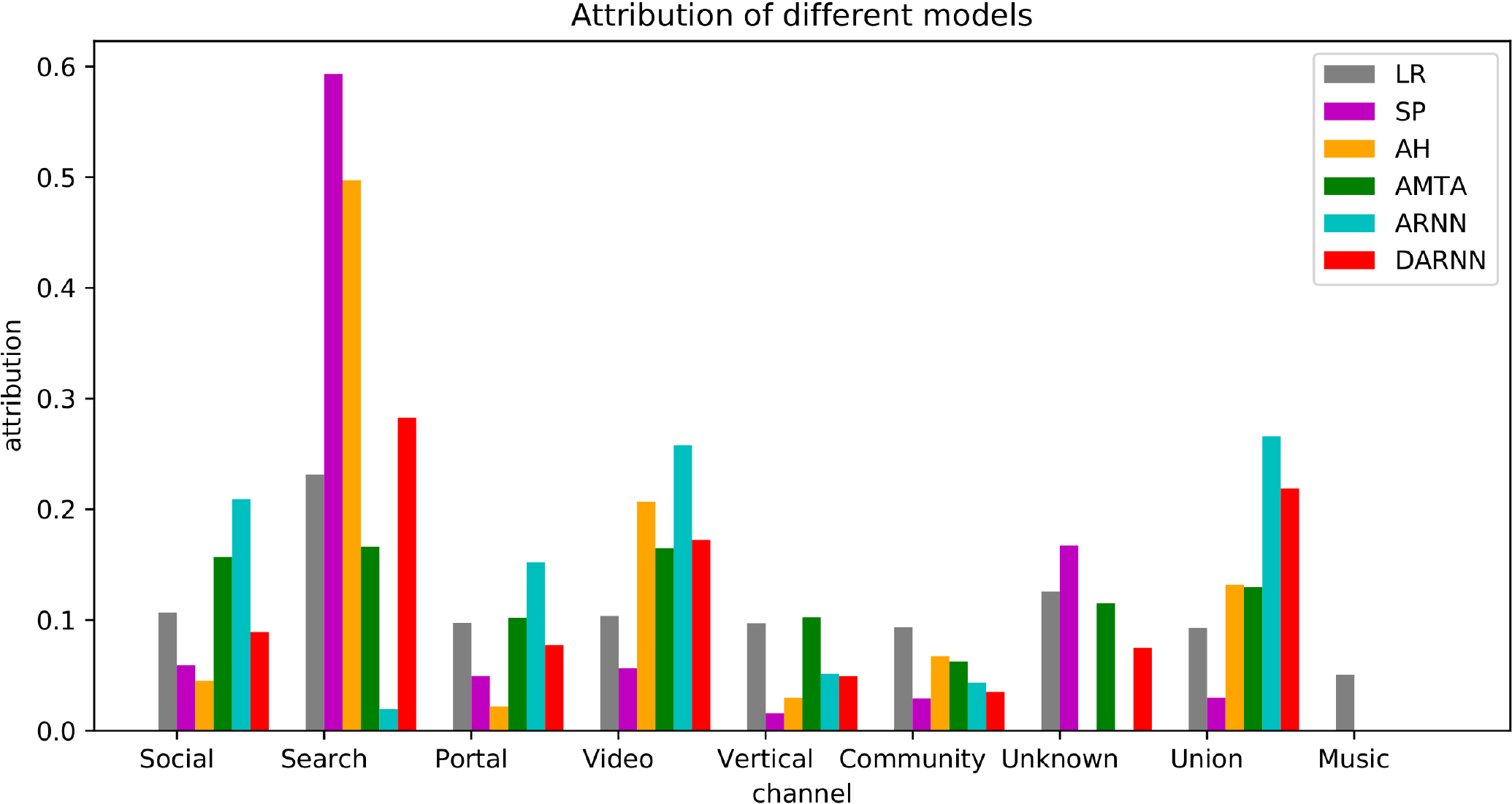}
	\caption{Attribution of different channels on Miaozhen.}\label{fig:attribution_miaozhen}
	\vspace{-10pt}
\end{figure}

The next analysis is based on the channel level attribution distribution.
We illustrate the calculated attribution credits of converted sequences over multiple channels in Miaozhen dataset as that in Figure~\ref{fig:attribution_miaozhen}.
The horizontal axis is the channel information varying from social media to music platforms.
Since there is no conversions on music channel, no credit has been assigned by the models except LR which takes the learned parameter coefficient of the channel feature for conversion attribution.
From the illustration we can find that
(i) LR, SP, AH and DARNN models assign the highest attribution credits to search channel, while ARNN and AMTA attribute the most onto video channel.
(ii) SP and AH assign relatively much higher credits on search channel, while the other models distribute attribution more smoothly.
(iii) Vertical and community channels have low credits while union channel has much higher attribution under attention-based models.
From these findings we find the significance of the replay evaluation for attribution guided budget allocation in the second stage of experiments, since the calculated conversion attribution credits over different channels vary from different models.

\begin{figure}[t]
	\centering
	\includegraphics[width=1.0\columnwidth]{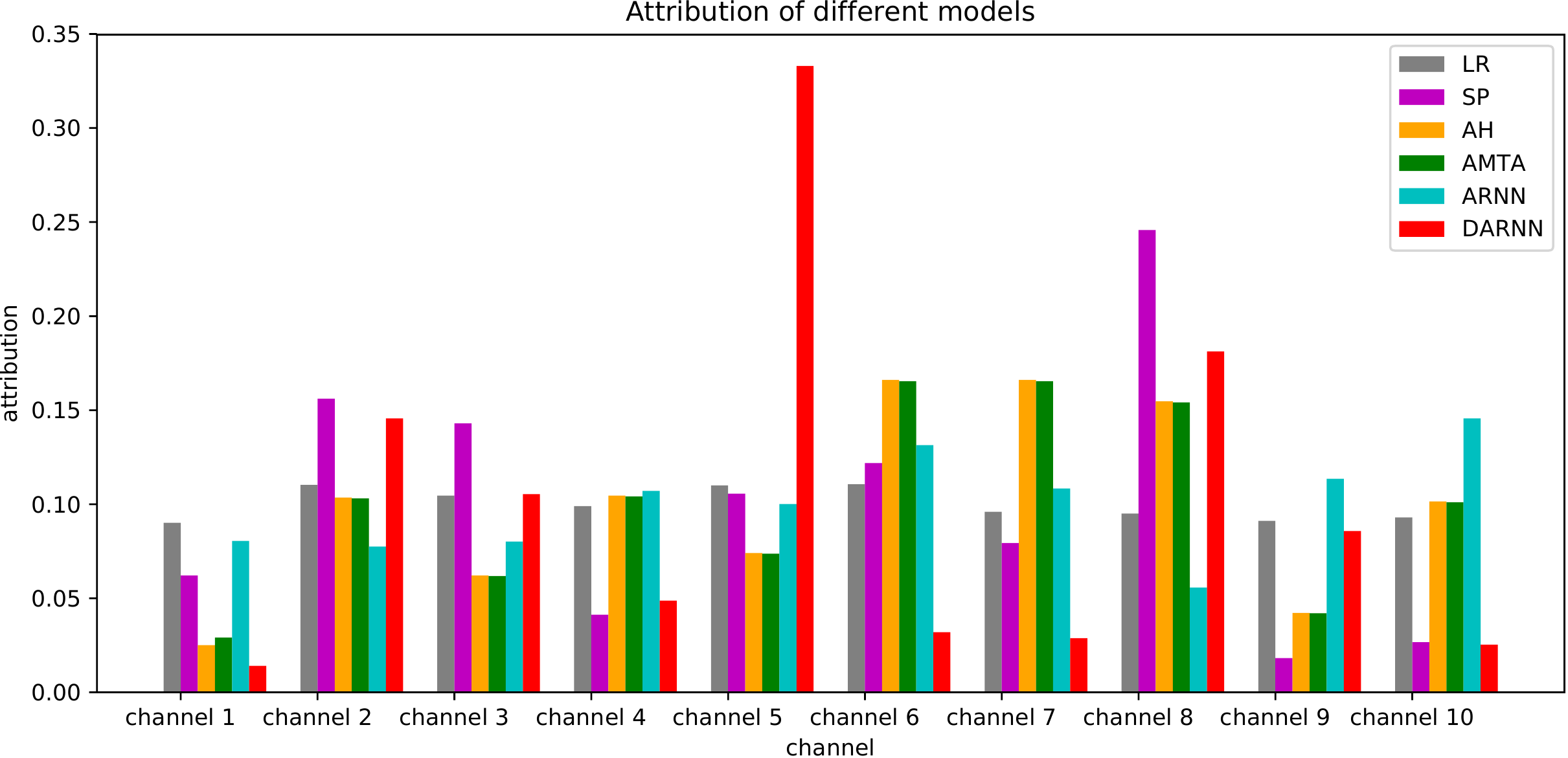}
	\caption{Attribution distribution over channels on Criteo.}\label{fig:attribution_criteo}
	\vspace{-5pt}
\end{figure}
\begin{figure}[t]
	\centering
	\includegraphics[width=1.0\columnwidth]{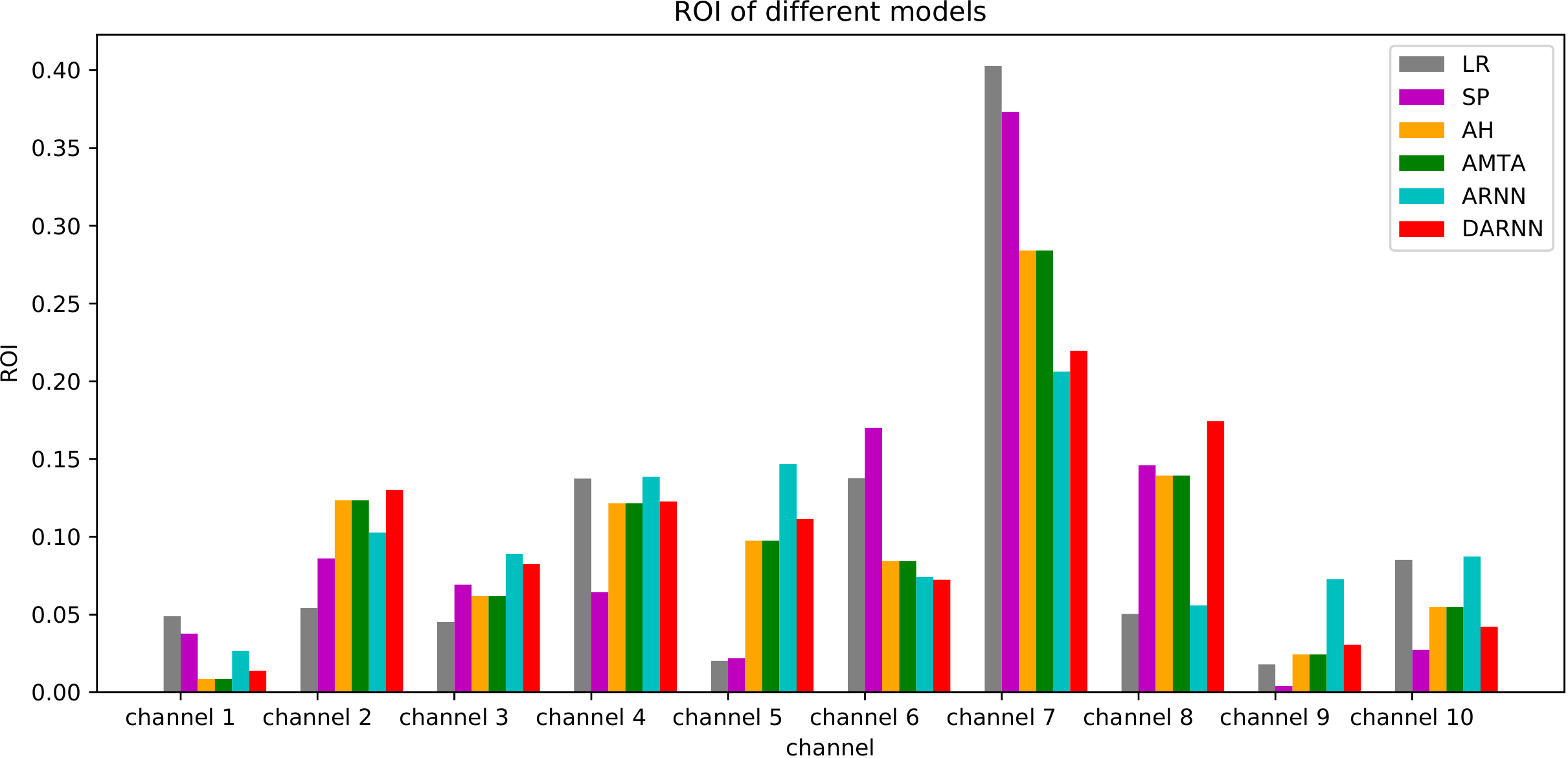}
	\caption{$\text{ROI}$ distribution over channels on Criteo.}\label{fig:ROI_criteo}
	\vspace{-10pt}
\end{figure}

In addition, we illustrate the attribution credits on Criteo dataset in Figure~\ref{fig:attribution_criteo} and the corresponding $\text{ROI}_{c_k}$ of each channel $c_k$ in Figure~\ref{fig:ROI_criteo} which is calculated according to Eq.~(\ref{eq:roi-calc}) under different models.
From channel level, our DARNN model assigns the highest credits onto channel 5.
However, the ROI calculation derives that all models allocate the most budget credits onto channel 7.
This is reasonable since the ROI is based on both channel level and touch point level information as that in Eq.~(\ref{eq:roi-calc}).

\begin{figure}[t]
	\centering
	\includegraphics[width=1.0\columnwidth]{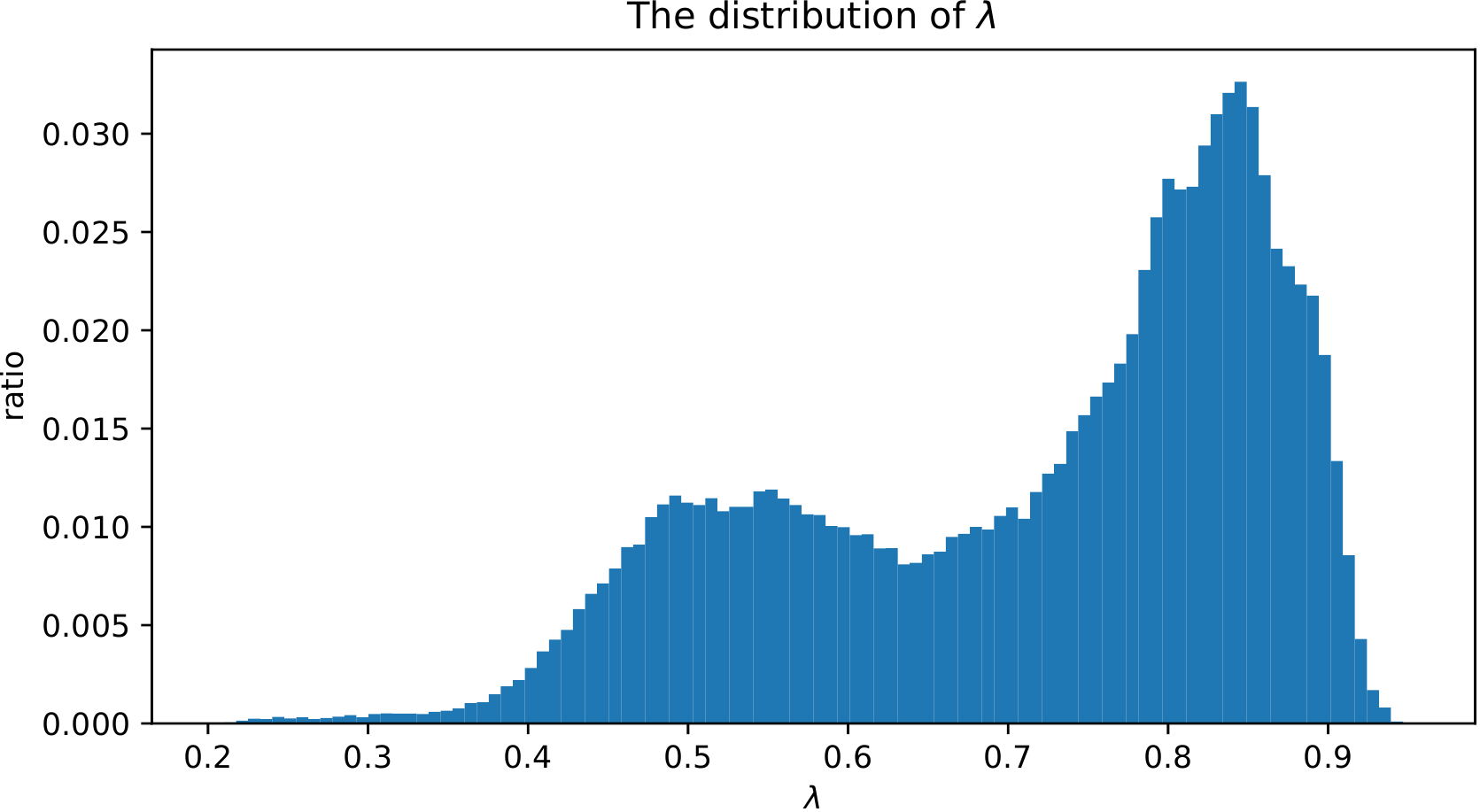}
	\caption{The distribution of $\lambda$ over Criteo dataset.}\label{fig:lambda-tuning}
	\vspace{-5pt}
\end{figure}

Finally, in Figure~\ref{fig:lambda-tuning}, we statistically visualize the value distribution of $\lambda$ which controls the impact from impression-level and click-level patterns, respectively.
As is calculated in Eq.~(\ref{eq:lambda-calculation}), note that, when $\lambda$ gets larger the click-level patterns get higher impact on the conversion attribution.
We may find from the figure that the click-level patterns relatively contribute more to the final conversion estimation, which reflects that the effectiveness of our dual-attention mechanism for different action pattern mining, as is described in Sec.~\ref{sec:attention}.
Generally speaking, the results illustrate the importance of combining both impression pattern and click pattern through dual-attention mechanism, especially that the click-level patterns contribute better under tight budget cases.

\section{Conclusion and Future Work}\label{sec:conclusion}
In this paper, we proposed a dual-attention recurrent neural network model for learning to assign conversion credits over the ad touch point sequences.
Our model not only captures sequential user patterns, but also pays attention to both impression-level and click-level user actions and derives an effective conversion attribution methodology.
The experiments show the significant improvement over the other state-of-the-art baselines.

One of the limitations of this work is that we have not taken the cost of ad impressions into account in the attention mechanism. It is of great interest to take the cost factor into modeling and improve the cost-effectiveness performance in the future as that in the works \cite{ren2016user,ren2017bidding} of real-time auction advertising.

\bibliographystyle{ACM-Reference-Format}
\balance
\bibliography{multi-attr}

%%% -*-BibTeX-*-
%%% Do NOT edit. File created by BibTeX with style
%%% ACM-Reference-Format-Journals [18-Jan-2012].

\begin{thebibliography}{37}

%%% ====================================================================
%%% NOTE TO THE USER: you can override these defaults by providing
%%% customized versions of any of these macros before the \bibliography
%%% command.  Each of them MUST provide its own final punctuation,
%%% except for \shownote{}, \showDOI{}, and \showURL{}.  The latter two
%%% do not use final punctuation, in order to avoid confusing it with
%%% the Web address.
%%%
%%% To suppress output of a particular field, define its macro to expand
%%% to an empty string, or better, \unskip, like this:
%%%
%%% \newcommand{\showDOI}[1]{\unskip}   % LaTeX syntax
%%%
%%% \def \showDOI #1{\unskip}           % plain TeX syntax
%%%
%%% ====================================================================

\ifx \showCODEN    \undefined \def \showCODEN     #1{\unskip}     \fi
\ifx \showDOI      \undefined \def \showDOI       #1{#1}\fi
\ifx \showISBNx    \undefined \def \showISBNx     #1{\unskip}     \fi
\ifx \showISBNxiii \undefined \def \showISBNxiii  #1{\unskip}     \fi
\ifx \showISSN     \undefined \def \showISSN      #1{\unskip}     \fi
\ifx \showLCCN     \undefined \def \showLCCN      #1{\unskip}     \fi
\ifx \shownote     \undefined \def \shownote      #1{#1}          \fi
\ifx \showarticletitle \undefined \def \showarticletitle #1{#1}   \fi
\ifx \showURL      \undefined \def \showURL       {\relax}        \fi
% The following commands are used for tagged output and should be
% invisible to TeX
\providecommand\bibfield[2]{#2}
\providecommand\bibinfo[2]{#2}
\providecommand\natexlab[1]{#1}
\providecommand\showeprint[2][]{arXiv:#2}

\bibitem[\protect\citeauthoryear{Agarwal, Ghosh, Wei, and You}{Agarwal
  et~al\mbox{.}}{2014}]%
        {agarwal2014budget}
\bibfield{author}{\bibinfo{person}{Deepak Agarwal}, \bibinfo{person}{Souvik
  Ghosh}, \bibinfo{person}{Kai Wei}, {and} \bibinfo{person}{Siyu You}.}
  \bibinfo{year}{2014}\natexlab{}.
\newblock \showarticletitle{Budget pacing for targeted online advertisements at
  LinkedIn}. In \bibinfo{booktitle}{\emph{KDD}}. ACM,
  \bibinfo{pages}{1613--1619}.
\newblock


\bibitem[\protect\citeauthoryear{Amin, Kearns, Key, and Schwaighofer}{Amin
  et~al\mbox{.}}{2012}]%
        {amin2012budget}
\bibfield{author}{\bibinfo{person}{Kareem Amin}, \bibinfo{person}{Michael
  Kearns}, \bibinfo{person}{Peter Key}, {and} \bibinfo{person}{Anton
  Schwaighofer}.} \bibinfo{year}{2012}\natexlab{}.
\newblock \showarticletitle{Budget optimization for sponsored search: Censored
  learning in {MDP}s}.
\newblock \bibinfo{journal}{\emph{UAI}} (\bibinfo{year}{2012}).
\newblock


\bibitem[\protect\citeauthoryear{Bahdanau, Cho, and Bengio}{Bahdanau
  et~al\mbox{.}}{2014}]%
        {bahdanau2014neural}
\bibfield{author}{\bibinfo{person}{Dzmitry Bahdanau},
  \bibinfo{person}{Kyunghyun Cho}, {and} \bibinfo{person}{Yoshua Bengio}.}
  \bibinfo{year}{2014}\natexlab{}.
\newblock \showarticletitle{Neural machine translation by jointly learning to
  align and translate}.
\newblock \bibinfo{journal}{\emph{arXiv preprint arXiv:1409.0473}}
  (\bibinfo{year}{2014}).
\newblock


\bibitem[\protect\citeauthoryear{Berman}{Berman}{2017}]%
        {berman2017beyond}
\bibfield{author}{\bibinfo{person}{Ron Berman}.}
  \bibinfo{year}{2017}\natexlab{}.
\newblock \showarticletitle{Beyond the last touch: Attribution in online
  advertising}.
\newblock  (\bibinfo{year}{2017}).
\newblock


\bibitem[\protect\citeauthoryear{Cao, Chen, Wang, Zhang, and Yu}{Cao
  et~al\mbox{.}}{2018}]%
        {cao2018neural}
\bibfield{author}{\bibinfo{person}{Xuezhi Cao}, \bibinfo{person}{Haokun Chen},
  \bibinfo{person}{Xuejian Wang}, \bibinfo{person}{Weinan Zhang}, {and}
  \bibinfo{person}{Yong Yu}.} \bibinfo{year}{2018}\natexlab{}.
\newblock \showarticletitle{Neural Link Prediction over Aligned Networks}. In
  \bibinfo{booktitle}{\emph{AAAI}}.
\newblock


\bibitem[\protect\citeauthoryear{Chandler-Pepelnjak}{Chandler-Pepelnjak}{2009}]%
        {chandler2009measuring}
\bibfield{author}{\bibinfo{person}{John Chandler-Pepelnjak}.}
  \bibinfo{year}{2009}\natexlab{}.
\newblock \showarticletitle{Measuring roi beyond the last ad}.
\newblock \bibinfo{journal}{\emph{Atlas Institute Digital Marketing Insight}}
  (\bibinfo{year}{2009}), \bibinfo{pages}{1--6}.
\newblock


\bibitem[\protect\citeauthoryear{Dalessandro, Perlich, Stitelman, and
  Provost}{Dalessandro et~al\mbox{.}}{2012}]%
        {dalessandro2012causally}
\bibfield{author}{\bibinfo{person}{Brian Dalessandro}, \bibinfo{person}{Claudia
  Perlich}, \bibinfo{person}{Ori Stitelman}, {and} \bibinfo{person}{Foster
  Provost}.} \bibinfo{year}{2012}\natexlab{}.
\newblock \showarticletitle{Causally motivated attribution for online
  advertising}. In \bibinfo{booktitle}{\emph{ADKDD}}. ACM, \bibinfo{pages}{7}.
\newblock


\bibitem[\protect\citeauthoryear{Diemert, Meynet, Galland, and
  Lefortier}{Diemert et~al\mbox{.}}{2017}]%
        {DiemertMeynet2017}
\bibfield{author}{\bibinfo{person}{Eustache Diemert}, \bibinfo{person}{Julien
  Meynet}, \bibinfo{person}{Pierre Galland}, {and} \bibinfo{person}{Damien
  Lefortier}.} \bibinfo{year}{2017}\natexlab{}.
\newblock \showarticletitle{Attribution Modeling Increases Efficiency of
  Bidding in Display Advertising}. In \bibinfo{booktitle}{\emph{ADKDD}}.
  \bibinfo{publisher}{ACM}.
\newblock


\bibitem[\protect\citeauthoryear{Gehring, Auli, Grangier, Yarats, and
  Dauphin}{Gehring et~al\mbox{.}}{2017}]%
        {gehring2017convolutional}
\bibfield{author}{\bibinfo{person}{Jonas Gehring}, \bibinfo{person}{Michael
  Auli}, \bibinfo{person}{David Grangier}, \bibinfo{person}{Denis Yarats},
  {and} \bibinfo{person}{Yann~N Dauphin}.} \bibinfo{year}{2017}\natexlab{}.
\newblock \showarticletitle{Convolutional sequence to sequence learning}.
\newblock \bibinfo{journal}{\emph{arXiv preprint arXiv:1705.03122}}
  (\bibinfo{year}{2017}).
\newblock


\bibitem[\protect\citeauthoryear{Geyik, Saxena, and Dasdan}{Geyik
  et~al\mbox{.}}{2014}]%
        {geyik2014multi}
\bibfield{author}{\bibinfo{person}{Sahin~Cem Geyik}, \bibinfo{person}{Abhishek
  Saxena}, {and} \bibinfo{person}{Ali Dasdan}.}
  \bibinfo{year}{2014}\natexlab{}.
\newblock \showarticletitle{Multi-touch attribution based budget allocation in
  online advertising}. In \bibinfo{booktitle}{\emph{ADKDD}}. ACM,
  \bibinfo{pages}{1--9}.
\newblock


\bibitem[\protect\citeauthoryear{Hochreiter and Schmidhuber}{Hochreiter and
  Schmidhuber}{1997}]%
        {hochreiter1997long}
\bibfield{author}{\bibinfo{person}{Sepp Hochreiter} {and}
  \bibinfo{person}{J{\"u}rgen Schmidhuber}.} \bibinfo{year}{1997}\natexlab{}.
\newblock \showarticletitle{Long short-term memory}.
\newblock \bibinfo{journal}{\emph{Neural computation}} (\bibinfo{year}{1997}).
\newblock


\bibitem[\protect\citeauthoryear{Ji and Wang}{Ji and Wang}{2017}]%
        {ji2017additional}
\bibfield{author}{\bibinfo{person}{Wendi Ji} {and} \bibinfo{person}{Xiaoling
  Wang}.} \bibinfo{year}{2017}\natexlab{}.
\newblock \showarticletitle{Additional Multi-Touch Attribution for Online
  Advertising}. In \bibinfo{booktitle}{\emph{AAAI}}.
\newblock


\bibitem[\protect\citeauthoryear{Ji, Wang, and Zhang}{Ji et~al\mbox{.}}{2016}]%
        {ji2016probabilistic}
\bibfield{author}{\bibinfo{person}{Wendi Ji}, \bibinfo{person}{Xiaoling Wang},
  {and} \bibinfo{person}{Dell Zhang}.} \bibinfo{year}{2016}\natexlab{}.
\newblock \showarticletitle{A probabilistic multi-touch attribution model for
  online advertising}. In \bibinfo{booktitle}{\emph{CIKM}}. ACM.
\newblock


\bibitem[\protect\citeauthoryear{Lee, Jalali, and Dasdan}{Lee
  et~al\mbox{.}}{2013}]%
        {lee2013real}
\bibfield{author}{\bibinfo{person}{Kuang-Chih Lee}, \bibinfo{person}{Ali
  Jalali}, {and} \bibinfo{person}{Ali Dasdan}.}
  \bibinfo{year}{2013}\natexlab{}.
\newblock \showarticletitle{Real time bid optimization with smooth budget
  delivery in online advertising}. In \bibinfo{booktitle}{\emph{ADKDD}}. ACM,
  \bibinfo{pages}{1}.
\newblock


\bibitem[\protect\citeauthoryear{Lee, Orten, Dasdan, and Li}{Lee
  et~al\mbox{.}}{2012}]%
        {lee2012estimating}
\bibfield{author}{\bibinfo{person}{Kuang-chih Lee}, \bibinfo{person}{Burkay
  Orten}, \bibinfo{person}{Ali Dasdan}, {and} \bibinfo{person}{Wentong Li}.}
  \bibinfo{year}{2012}\natexlab{}.
\newblock \showarticletitle{Estimating conversion rate in display advertising
  from past performance data}. In \bibinfo{booktitle}{\emph{KDD}}. ACM,
  \bibinfo{pages}{768--776}.
\newblock


\bibitem[\protect\citeauthoryear{Ma, Zhao, Huang, Wang, Hu, Zhu, and Gai}{Ma
  et~al\mbox{.}}{2018}]%
        {ma2018entire}
\bibfield{author}{\bibinfo{person}{Xiao Ma}, \bibinfo{person}{Liqin Zhao},
  \bibinfo{person}{Guan Huang}, \bibinfo{person}{Zhi Wang},
  \bibinfo{person}{Zelin Hu}, \bibinfo{person}{Xiaoqiang Zhu}, {and}
  \bibinfo{person}{Kun Gai}.} \bibinfo{year}{2018}\natexlab{}.
\newblock \showarticletitle{Entire Space Multi-Task Model: An Effective
  Approach for Estimating Post-Click Conversion Rate}.
\newblock \bibinfo{journal}{\emph{SIGIR}} (\bibinfo{year}{2018}).
\newblock


\bibitem[\protect\citeauthoryear{McMahan, Holt, Sculley, Young, Ebner, Grady,
  Nie, Phillips, Davydov, Golovin, et~al\mbox{.}}{McMahan
  et~al\mbox{.}}{2013}]%
        {mcmahan2013ad}
\bibfield{author}{\bibinfo{person}{H~Brendan McMahan}, \bibinfo{person}{Gary
  Holt}, \bibinfo{person}{David Sculley}, \bibinfo{person}{Michael Young},
  \bibinfo{person}{Dietmar Ebner}, \bibinfo{person}{Julian Grady},
  \bibinfo{person}{Lan Nie}, \bibinfo{person}{Todd Phillips},
  \bibinfo{person}{Eugene Davydov}, \bibinfo{person}{Daniel Golovin},
  {et~al\mbox{.}}} \bibinfo{year}{2013}\natexlab{}.
\newblock \showarticletitle{Ad click prediction: a view from the trenches}. In
  \bibinfo{booktitle}{\emph{KDD}}.
\newblock


\bibitem[\protect\citeauthoryear{Perlich, Dalessandro, Hook, Stitelman, Raeder,
  and Provost}{Perlich et~al\mbox{.}}{2012}]%
        {perlich2012bid}
\bibfield{author}{\bibinfo{person}{Claudia Perlich}, \bibinfo{person}{Brian
  Dalessandro}, \bibinfo{person}{Rod Hook}, \bibinfo{person}{Ori Stitelman},
  \bibinfo{person}{Troy Raeder}, {and} \bibinfo{person}{Foster Provost}.}
  \bibinfo{year}{2012}\natexlab{}.
\newblock \showarticletitle{Bid optimizing and inventory scoring in targeted
  online advertising}. In \bibinfo{booktitle}{\emph{KDD}}. ACM,
  \bibinfo{pages}{804--812}.
\newblock


\bibitem[\protect\citeauthoryear{Qin, Song, Cheng, Cheng, Jiang, and
  Cottrell}{Qin et~al\mbox{.}}{2017}]%
        {qin2017dual}
\bibfield{author}{\bibinfo{person}{Yao Qin}, \bibinfo{person}{Dongjin Song},
  \bibinfo{person}{Haifeng Cheng}, \bibinfo{person}{Wei Cheng},
  \bibinfo{person}{Guofei Jiang}, {and} \bibinfo{person}{Garrison Cottrell}.}
  \bibinfo{year}{2017}\natexlab{}.
\newblock \showarticletitle{A dual-stage attention-based recurrent neural
  network for time series prediction}.
\newblock \bibinfo{journal}{\emph{arXiv preprint arXiv:1704.02971}}
  (\bibinfo{year}{2017}).
\newblock


\bibitem[\protect\citeauthoryear{Qu, Cai, Ren, Zhang, Yu, Wen, and Wang}{Qu
  et~al\mbox{.}}{2016}]%
        {qu2016product}
\bibfield{author}{\bibinfo{person}{Yanru Qu}, \bibinfo{person}{Han Cai},
  \bibinfo{person}{Kan Ren}, \bibinfo{person}{Weinan Zhang},
  \bibinfo{person}{Yong Yu}, \bibinfo{person}{Ying Wen}, {and}
  \bibinfo{person}{Jun Wang}.} \bibinfo{year}{2016}\natexlab{}.
\newblock \showarticletitle{Product-based neural networks for user response
  prediction}. In \bibinfo{booktitle}{\emph{ICDM}}.
\newblock


\bibitem[\protect\citeauthoryear{Ren, Zhang, Chang, Rong, Yu, and Wang}{Ren
  et~al\mbox{.}}{2018}]%
        {ren2017bidding}
\bibfield{author}{\bibinfo{person}{Kan Ren}, \bibinfo{person}{Weinan Zhang},
  \bibinfo{person}{Ke Chang}, \bibinfo{person}{Yifei Rong},
  \bibinfo{person}{Yong Yu}, {and} \bibinfo{person}{Jun Wang}.}
  \bibinfo{year}{2018}\natexlab{}.
\newblock \showarticletitle{Bidding Machine: Learning to Bid for Directly
  Optimizing Profits in Display Advertising}.
\newblock \bibinfo{journal}{\emph{TKDE}} (\bibinfo{year}{2018}).
\newblock


\bibitem[\protect\citeauthoryear{Ren, Zhang, Rong, Zhang, Yu, and Wang}{Ren
  et~al\mbox{.}}{2016}]%
        {ren2016user}
\bibfield{author}{\bibinfo{person}{Kan Ren}, \bibinfo{person}{Weinan Zhang},
  \bibinfo{person}{Yifei Rong}, \bibinfo{person}{Haifeng Zhang},
  \bibinfo{person}{Yong Yu}, {and} \bibinfo{person}{Jun Wang}.}
  \bibinfo{year}{2016}\natexlab{}.
\newblock \showarticletitle{User response learning for directly optimizing
  campaign performance in display advertising}. In
  \bibinfo{booktitle}{\emph{CIKM}}. ACM, \bibinfo{pages}{679--688}.
\newblock


\bibitem[\protect\citeauthoryear{Rendle, Freudenthaler, and
  Schmidt-Thieme}{Rendle et~al\mbox{.}}{2010}]%
        {rendle2010factorizing}
\bibfield{author}{\bibinfo{person}{Steffen Rendle}, \bibinfo{person}{Christoph
  Freudenthaler}, {and} \bibinfo{person}{Lars Schmidt-Thieme}.}
  \bibinfo{year}{2010}\natexlab{}.
\newblock \showarticletitle{Factorizing personalized markov chains for
  next-basket recommendation}. In \bibinfo{booktitle}{\emph{WWW}}.
\newblock


\bibitem[\protect\citeauthoryear{Shao and Li}{Shao and Li}{2011}]%
        {shao2011data}
\bibfield{author}{\bibinfo{person}{Xuhui Shao} {and} \bibinfo{person}{Lexin
  Li}.} \bibinfo{year}{2011}\natexlab{}.
\newblock \showarticletitle{Data-driven multi-touch attribution models}. In
  \bibinfo{booktitle}{\emph{KDD}}. ACM, \bibinfo{pages}{258--264}.
\newblock


\bibitem[\protect\citeauthoryear{Sinha, Saini, and Anadhavelu}{Sinha
  et~al\mbox{.}}{2014}]%
        {sinha2014estimating}
\bibfield{author}{\bibinfo{person}{Ritwik Sinha}, \bibinfo{person}{Shiv Saini},
  {and} \bibinfo{person}{N Anadhavelu}.} \bibinfo{year}{2014}\natexlab{}.
\newblock \showarticletitle{Estimating the incremental effects of interactions
  for marketing attribution}. In \bibinfo{booktitle}{\emph{Behavior, Economic
  and Social Computing (BESC), 2014 International Conference on}}. IEEE,
  \bibinfo{pages}{1--6}.
\newblock


\bibitem[\protect\citeauthoryear{Song, Xiao, Wu, Wu, Zhang, Zhang, and
  Zhu}{Song et~al\mbox{.}}{2017}]%
        {song2017hierarchical}
\bibfield{author}{\bibinfo{person}{Jun Song}, \bibinfo{person}{Jun Xiao},
  \bibinfo{person}{Fei Wu}, \bibinfo{person}{Haishan Wu}, \bibinfo{person}{Tong
  Zhang}, \bibinfo{person}{Zhongfei Zhang}, {and} \bibinfo{person}{Wenwu Zhu}.}
  \bibinfo{year}{2017}\natexlab{}.
\newblock \showarticletitle{Hierarchical Contextual Attention Recurrent Neural
  Network for Map Query Suggestion}.
\newblock \bibinfo{journal}{\emph{TKDE}} (\bibinfo{year}{2017}).
\newblock


\bibitem[\protect\citeauthoryear{Wang, Zhang, and Yuan}{Wang
  et~al\mbox{.}}{2017b}]%
        {wang2017display}
\bibfield{author}{\bibinfo{person}{Jun Wang}, \bibinfo{person}{Weinan Zhang},
  {and} \bibinfo{person}{Shuai Yuan}.} \bibinfo{year}{2017}\natexlab{b}.
\newblock \showarticletitle{Display Advertising with Real-Time Bidding (RTB)
  and Behavioural Targeting}.
\newblock \bibinfo{journal}{\emph{Now Publisher}} (\bibinfo{year}{2017}).
\newblock


\bibitem[\protect\citeauthoryear{Wang, Yu, Ren, Tao, Zhang, Yu, and Wang}{Wang
  et~al\mbox{.}}{2017a}]%
        {wang2017dynamic}
\bibfield{author}{\bibinfo{person}{Xuejian Wang}, \bibinfo{person}{Lantao Yu},
  \bibinfo{person}{Kan Ren}, \bibinfo{person}{Guanyu Tao},
  \bibinfo{person}{Weinan Zhang}, \bibinfo{person}{Yong Yu}, {and}
  \bibinfo{person}{Jun Wang}.} \bibinfo{year}{2017}\natexlab{a}.
\newblock \showarticletitle{Dynamic attention deep model for article
  recommendation by learning human editors' demonstration}. In
  \bibinfo{booktitle}{\emph{KDD}}. ACM.
\newblock


\bibitem[\protect\citeauthoryear{Xu, Shao, Ma, Lee, Qi, and Lu}{Xu
  et~al\mbox{.}}{2016}]%
        {xu2016lift}
\bibfield{author}{\bibinfo{person}{Jian Xu}, \bibinfo{person}{Xuhui Shao},
  \bibinfo{person}{Jianjie Ma}, \bibinfo{person}{Kuang-chih Lee},
  \bibinfo{person}{Hang Qi}, {and} \bibinfo{person}{Quan Lu}.}
  \bibinfo{year}{2016}\natexlab{}.
\newblock \showarticletitle{Lift-based bidding in ad selection}. In
  \bibinfo{booktitle}{\emph{AAAI}}.
\newblock


\bibitem[\protect\citeauthoryear{Xu, Duan, and Whinston}{Xu
  et~al\mbox{.}}{2014}]%
        {xu2014path}
\bibfield{author}{\bibinfo{person}{Lizhen Xu}, \bibinfo{person}{Jason~A Duan},
  {and} \bibinfo{person}{Andrew Whinston}.} \bibinfo{year}{2014}\natexlab{}.
\newblock \showarticletitle{Path to purchase: A mutually exciting point process
  model for online advertising and conversion}.
\newblock \bibinfo{journal}{\emph{Management Science}} \bibinfo{volume}{60},
  \bibinfo{number}{6} (\bibinfo{year}{2014}), \bibinfo{pages}{1392--1412}.
\newblock


\bibitem[\protect\citeauthoryear{Yan, Zhang, Zha, Gong, Sun, Huang, Chu, and
  Yang}{Yan et~al\mbox{.}}{2015}]%
        {yan2015machine}
\bibfield{author}{\bibinfo{person}{Junchi Yan}, \bibinfo{person}{Chao Zhang},
  \bibinfo{person}{Hongyuan Zha}, \bibinfo{person}{Min Gong},
  \bibinfo{person}{Changhua Sun}, \bibinfo{person}{Jin Huang},
  \bibinfo{person}{Stephen Chu}, {and} \bibinfo{person}{Xiaokang Yang}.}
  \bibinfo{year}{2015}\natexlab{}.
\newblock \showarticletitle{On machine learning towards predictive sales
  pipeline analytics}. In \bibinfo{booktitle}{\emph{AAAI}}.
\newblock


\bibitem[\protect\citeauthoryear{Yuan}{Yuan}{2015}]%
        {yuan2015supply}
\bibfield{author}{\bibinfo{person}{Shuai Yuan}.}
  \bibinfo{year}{2015}\natexlab{}.
\newblock \emph{\bibinfo{title}{Supply side optimisation in online display
  advertising (Chapter 3)}}.
\newblock \bibinfo{thesistype}{Ph.D. Dissertation}. \bibinfo{school}{UCL
  (University College London)}.
\newblock


\bibitem[\protect\citeauthoryear{Zhai, Chang, Zhang, and Zhang}{Zhai
  et~al\mbox{.}}{2016}]%
        {zhai2016deepintent}
\bibfield{author}{\bibinfo{person}{Shuangfei Zhai}, \bibinfo{person}{Keng-hao
  Chang}, \bibinfo{person}{Ruofei Zhang}, {and} \bibinfo{person}{Zhongfei~Mark
  Zhang}.} \bibinfo{year}{2016}\natexlab{}.
\newblock \showarticletitle{Deepintent: Learning attentions for online
  advertising with recurrent neural networks}. In
  \bibinfo{booktitle}{\emph{KDD}}. ACM.
\newblock


\bibitem[\protect\citeauthoryear{Zhang, Du, and Wang}{Zhang
  et~al\mbox{.}}{2016}]%
        {zhang2016deep}
\bibfield{author}{\bibinfo{person}{Weinan Zhang}, \bibinfo{person}{Tianming
  Du}, {and} \bibinfo{person}{Jun Wang}.} \bibinfo{year}{2016}\natexlab{}.
\newblock \showarticletitle{Deep Learning over Multi-field Categorical Data: A
  Case Study on User Response Prediction}.
\newblock \bibinfo{journal}{\emph{ECIR}} (\bibinfo{year}{2016}).
\newblock


\bibitem[\protect\citeauthoryear{Zhang, Yuan, and Wang}{Zhang
  et~al\mbox{.}}{2014c}]%
        {zhang2014optimal}
\bibfield{author}{\bibinfo{person}{Weinan Zhang}, \bibinfo{person}{Shuai Yuan},
  {and} \bibinfo{person}{Jun Wang}.} \bibinfo{year}{2014}\natexlab{c}.
\newblock \showarticletitle{Optimal real-time bidding for display advertising}.
  In \bibinfo{booktitle}{\emph{KDD}}. ACM, \bibinfo{pages}{1077--1086}.
\newblock


\bibitem[\protect\citeauthoryear{Zhang, Dai, Xu, Feng, Wang, Bian, Wang, and
  Liu}{Zhang et~al\mbox{.}}{2014a}]%
        {zhang2014sequential}
\bibfield{author}{\bibinfo{person}{Yuyu Zhang}, \bibinfo{person}{Hanjun Dai},
  \bibinfo{person}{Chang Xu}, \bibinfo{person}{Jun Feng},
  \bibinfo{person}{Taifeng Wang}, \bibinfo{person}{Jiang Bian},
  \bibinfo{person}{Bin Wang}, {and} \bibinfo{person}{Tie-Yan Liu}.}
  \bibinfo{year}{2014}\natexlab{a}.
\newblock \showarticletitle{Sequential click prediction for sponsored search
  with recurrent neural networks}.
\newblock \bibinfo{journal}{\emph{arXiv preprint arXiv:1404.5772}}
  (\bibinfo{year}{2014}).
\newblock


\bibitem[\protect\citeauthoryear{Zhang, Wei, and Ren}{Zhang
  et~al\mbox{.}}{2014b}]%
        {zhang2014multi}
\bibfield{author}{\bibinfo{person}{Ya Zhang}, \bibinfo{person}{Yi Wei}, {and}
  \bibinfo{person}{Jianbiao Ren}.} \bibinfo{year}{2014}\natexlab{b}.
\newblock \showarticletitle{Multi-touch attribution in online advertising with
  survival theory}. In \bibinfo{booktitle}{\emph{ICDM}}. IEEE,
  \bibinfo{pages}{687--696}.
\newblock


\end{thebibliography}

\end{document}